\documentclass[acmsmall]{acmart}
\AtBeginDocument{%
  \providecommand\BibTeX{{%
    \normalfont B\kern-0.5em{\scshape i\kern-0.25em b}\kern-0.8em\TeX}}}

\usepackage{xspace}
\usepackage{caption}
\usepackage{subcaption}
\usepackage{multirow}

\newcommand{\ie}{\emph{i.e.,}\xspace}
\newcommand{\eg}{\emph{e.g.,}\xspace}

\newtheorem{observation} {Finding}
\newcommand{\dataset}{MovieLens1518\xspace}

\copyrightyear{2024} 
\acmYear{2024} 
\setcopyright{rightsretained}
\acmDOI{xxxxxx.xxxxxx}

\acmJournal{TOIS}
\acmVolume{1}
\acmNumber{1}
\acmArticle{111}
\acmMonth{3}

\begin{document}

\title{Our Model Achieves Excellent Performance on MovieLens: What Does It Mean?}

\author{Yu-chen Fan}
 \email{S220088@e.ntu.edu.sg}
 \affiliation{%
   \institution{Nanyang Technological University}
   \country{Singapore}
   \postcode{639798}
 }
\author{Yitong Ji}
 \email{S190004@e.ntu.edu.sg}
 \affiliation{%
   \institution{Nanyang Technological University}
   \country{Singapore}
   \postcode{639798}
 }

\author{Jie Zhang}
\email{zhangj@ntu.edu.sg}
\affiliation{%
  \institution{Nanyang Technological University}
  \country{Singapore}
  \postcode{639798}
}
\orcid{}

\author{Aixin Sun}
\authornote{Corresponding author.}
\email{axsun@ntu.edu.sg}
\affiliation{%
  \institution{Nanyang Technological University}
  \country{Singapore}
  \postcode{639798}
}
\orcid{0000-0003-0764-4258}

\begin{abstract}
A typical benchmark dataset for recommender system (RecSys) evaluation consists of user-item interactions generated on a platform within a time period. The interaction generation mechanism partially explains why a user interacts with (\eg like, purchase, rate) an item, and the context of when a particular interaction happened. In this study, we conduct a meticulous analysis of the MovieLens dataset and explain the potential impact of using the dataset for evaluating recommendation algorithms. We make a few main findings from our analysis. First, there are significant differences in user interactions at the different stages when a user interacts with the MovieLens platform. The early interactions largely define the user portrait which affect the subsequent interactions. Second, user interactions are highly affected by the candidate movies that are recommended by the platform's internal recommendation algorithm(s).  Third, changing the order of user interactions makes it more difficult for sequential algorithms to capture the progressive interaction process. We further discuss the discrepancy between the interaction generation mechanism that is employed by the MovieLens system and that of typical real-world recommendation scenarios. That is, the MovieLens dataset records $\langle user - MovieLens\rangle$ interactions, but not $\langle user - movie \rangle$ interactions. All research papers using the MovieLens dataset model the  $\langle user - MovieLens\rangle$ rather than the $\langle user - movie\rangle$ interactions, making their results less generalizable to many practical recommendation scenarios in real-world settings. In summary, the MovieLens platform demonstrates an efficient and effective way of collecting user preferences to address cold-starts. However, \textit{models that achieve excellent recommendation accuracy on the MovieLens dataset may not demonstrate superior performance in practice}, for at least two kinds of differences: (i) the differences in the contexts of user-item interaction generation, and (ii) the differences in user knowledge about the item collections.  
While results on MovieLens can be useful as a reference, they should not be solely relied upon as the primary justification for the effectiveness of a recommendation system model. 
\end{abstract}

\ccsdesc[500]{Information systems~Recommender systems}
\ccsdesc[300]{Information systems~Evaluation of retrieval results}


\keywords{Recommendation Evaluation, MovieLens, Data Analysis}
\maketitle

\section{Introduction}
\label{sec:intro}

The rapid growth of digital platforms and data-driven services in recent years has given rise to a thriving recommendation system landscape, fueled by the continuous development and refinement of algorithms to meet the diverse needs of users. With an ever-increasing volume of data at their disposal, these algorithms leverage sophisticated techniques, such as collaborative filtering~\cite{DBLP:journals/tois/DeshpandeK04,DBLP:conf/recsys/CremonesiKT10}, sequence-aware recommendation~\cite{DBLP:conf/icdm/KangM18,DBLP:journals/corr/HidasiKBT15}, and deep learning~\cite{liang2018variational,wu2016collaborative,he2017neural}, to provide users with highly personalized and relevant content.

However, alongside the undeniable benefits, concerns surrounding the proper \textit{evaluation} of recommendation algorithms remain vital issues to address as the technology continues to advance. The evaluation of recommendation algorithms faces a multitude of challenges, primarily arising from the difficulties in reproducing results and the lack of uniform standards for comparison. For reproducibility, \citet{ferrari2019we} report that many algorithms have difficulty reproducing the results they show in their original papers, and some algorithms can be defeated by simpler algorithms. The absence of a unified set of evaluation metrics and methodologies compounds these challenges, as it hampers the ability to conduct fair and accurate comparisons between different algorithms. For example, to speed up assessment, sampled metrics evaluate algorithms based on selected negative samples. However, the results obtained from sampled metrics may be misleading, as it is inconsistent with the results of full rank~\cite{krichene2022sampled}. Accordingly, the  community has witnessed a number of efforts~\cite{sun2020we,zhao2022revisiting} to establish standardized evaluation protocols and to ensure that advancements in the field are grounded in reliable and reproducible research. However, standardized evaluation protocols are meaningful only if \textit{the datasets used for the evaluations are meaningful and representative}. In other words, a good understanding of each benchmark dataset used for evaluation is critical, \eg to what extent a benchmark dataset represents a recommendation scenario, and what scenario. If a dataset well represents a practical recommendation setting, then we can expect the algorithm stands for a good chance of achieving similar performance in practice as on the dataset. 

Indeed, researchers recently start to focus on an often overlooked aspect of the recommender system evaluation: \textit{the benchmark dataset}. Specifically, the characteristics of a dataset can significantly impact the performance of various algorithms obtained on the dataset~\cite{chin2022datasets}.  \citet{schnabel2016recommendations} explore the selection bias in implicit feedback datasets, which arises from users' selective exposure to items.  Our study expands on previous studies on dataset analysis and attempts to explore a  different perspective: \textit{\textbf {to what extent do we understand the user-item interaction generation mechanisms of a dataset?}} The findings made along this perspective help us to better understand model performance on the dataset, and make better expectations of model performance in practical settings. 

To this end, we have conducted a comprehensive and in-depth case study on the MovieLens dataset, arguably the most widely used benchmark dataset in the recommendation research community~\cite{sun2020we, chin2022datasets,JannachZGG12,klimashevskaia2023survey}. Our analysis is conducted in three steps.

First, we conduct a literature survey, combined with our own user experiences interacting with the MovieLens platform, to gain insights into the user-item interaction generation mechanisms of the platform. The movie ratings are collected through a progressive interaction process. In particular, interactions for most users on MovieLens are collected within a very short time period (\eg within a single day for about half of all users). The interactions from every user are much affected by the candidate movies listed on the recommended web pages at different stages during this progressive interaction collection process. The pool of candidate movies in the recommended page expands along the deepening of user interaction, thus causing differences in the recorded user interactions in different stages. Among them, the initial set of user interactions largely defines a user portrait, which affects the types of movies to be rated by this user.

Second, based on our comprehension of the user-item interaction generation mechanism, we design experiments to capture the potential impact of the data characteristics to the performance of recommendation algorithms. Specifically, we conduct ablation experiments on the raw data by masking out interactions at different stages of user interactions, and observe the performance changes of recommendation algorithms. 
The performance drop of the algorithm is most pronounced when we mask out the interactions that are closer to the predicted interaction. Following early studies, we also perform data shuffling on the rating sequences. The signal of causality between interactions introduced by the internal data collection mechanism of MovieLens is the main reason why sequential recommendation algorithms perform well on this dataset.

Finally, based on the findings, we discuss the discrepancy between the user-item interaction generation mechanisms of the MovieLens dataset and that of practical scenarios. Specifically, we discuss two kinds of differences. The first type of difference is the \textit{\textbf{contexts of user-item interaction generation}}. On a typical practical recommendation platform, a user may consider many factors before an interaction happens, \eg the monetary cost for purchasing a product or the time cost of watching a video. However, on MovieLens, the rating interaction is a record of what happened in the past \eg watched a movie at somewhere on someday. That's why nearly half of the users are able to record a good number of ratings within a single day on MovieLens. The decision-making process is very different.  The second type of difference is \textit{\textbf{the user knowledge about the item collections}}. We would expect most users to have a reasonable understanding of different movie genres like action, comedy, and romance, as well as a few key attributes like directors and actors. The same applies to recommendation settings for books and music. However, it is unreasonable to assume that most users have a good understanding of all kinds of products available on Amazon or eBay. The information gap between users and the item collection may affect user behavior. Accordingly, models that achieve good performance on MovieLens may not show the same under a recommendation scenario where users have limited knowledge about the item collections.

In summary, our work highlights the non-negligible role of datasets for model generalization and evaluation. To the best of our knowledge, this is the first attempt to design evaluation experiments and explain algorithm performance from the perspective of the interaction generation mechanism of a dataset. Based on our findings, we argue that the context of decision makings for users when interacting with the MovieLens platform could be very different from many other kinds of decision makings when users interact with a recommender platform. As a model is trained to learn the underlying patterns from a given dataset, the recommenders that achieve excellent performance on the MovieLens dataset may not show a competitive advantage in many real-world settings. In other words, while results on MovieLens can be useful as a reference, they should not be solely relied upon as the primary justification for the effectiveness of a recommendation system model.

\section{The MovieLens Interaction Generation Mechanism}
\label{sec:dataCollection}
In this section, we first brief the process of collecting user-movie ratings on the MovieLens platform. Based on our understanding of the data collection process, we curate a new dataset named \dataset, and provide an initial analysis on the dataset.

\subsection{Data Collection Mechanism}
\label{ssec:collectionInteractions}

The MovieLens dataset was initially introduced by the GroupLens research team in 1998. The interactions (\ie user ratings on movies) originate from the MovieLens recommendation platform, which aims to offer personalized movie recommendations based on users' preferences and viewing history. Over the years, the MovieLens platform has gone through a few versions to incorporate new features, improve user experience, and adapt to advancements in technology and recommender systems research. In 2015, the team published a paper~\cite{harper2015movielens} detailing the key transformations the platform underwent, which subsequently impact the resultant datasets. Specifically, the platform's intrinsic recommendation mechanism comprises two primary components: \textit{preference elicitation}~\cite{chang2015using}, and \textit{recommendation algorithm selection}~\cite{ekstrand2015letting}. To ensure the consistency in user-item interaction collection context, hence the sampled data used in our experiments, we focus on the latest version (Version 4) of the MovieLens platform.

\begin{figure*}[t]
    \centering
    \includegraphics[trim = 3cm 7cm 3cm 10cm, clip, width = 1.0\linewidth]{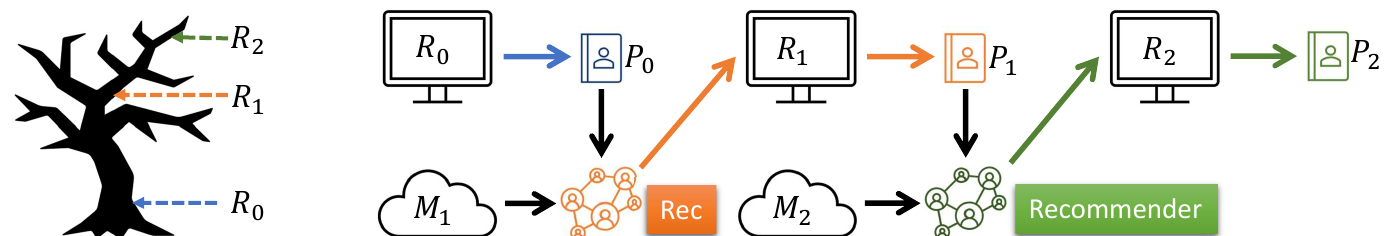}
    \caption{Illustration of the three stages ($R_0$, $R_1$, and $R_2$) when a user interacts with MovieLens. From $R_0$ to $R_3$, the search space of candidate movies becomes larger and the movies recommended become more personalized, from a common root to various branches in a tree representation (the figure on the left-hand side).  In the figure on the right-hand side, $M_1$ and $M_2$ represent the two pools of candidate movies to generate the list of recommended movies for users to rate, in stages $R_1$ and $R_2$ respectively. $P_0$ is the group-based user preference. $P_1$ and $P_2$ are the item-based preferences (\ie movie ratings) by different internal recommendation algorithms on MovieLens with candidate pools $M_1$ and $M_2$ respectively. }
    \label{fig:process}
\end{figure*}

The core user feedback on MovieLens is user ratings of movies that are recommended by this platform. The platform updates a user's portrait by capturing the user's ratings on a list of candidate movies in a web page. Then, the platform generates a new list of candidate movies in the next page for the user to rate, and the process repeats. The interaction generation mechanism between a user and the platform can be divided into three main stages: \textit{group-based preference elicitation} ($R_0$), \textit{item-based preference elicitation} ($R_1$), and \textit{recommendation with preferred algorithm} ($R_2$). Figure~\ref{fig:process} provides an illustration of the three stages.

\begin{figure*}[t]
    \centering
    \begin{subfigure}{0.85\textwidth}
    \centering
    \includegraphics[width = 0.75\linewidth]{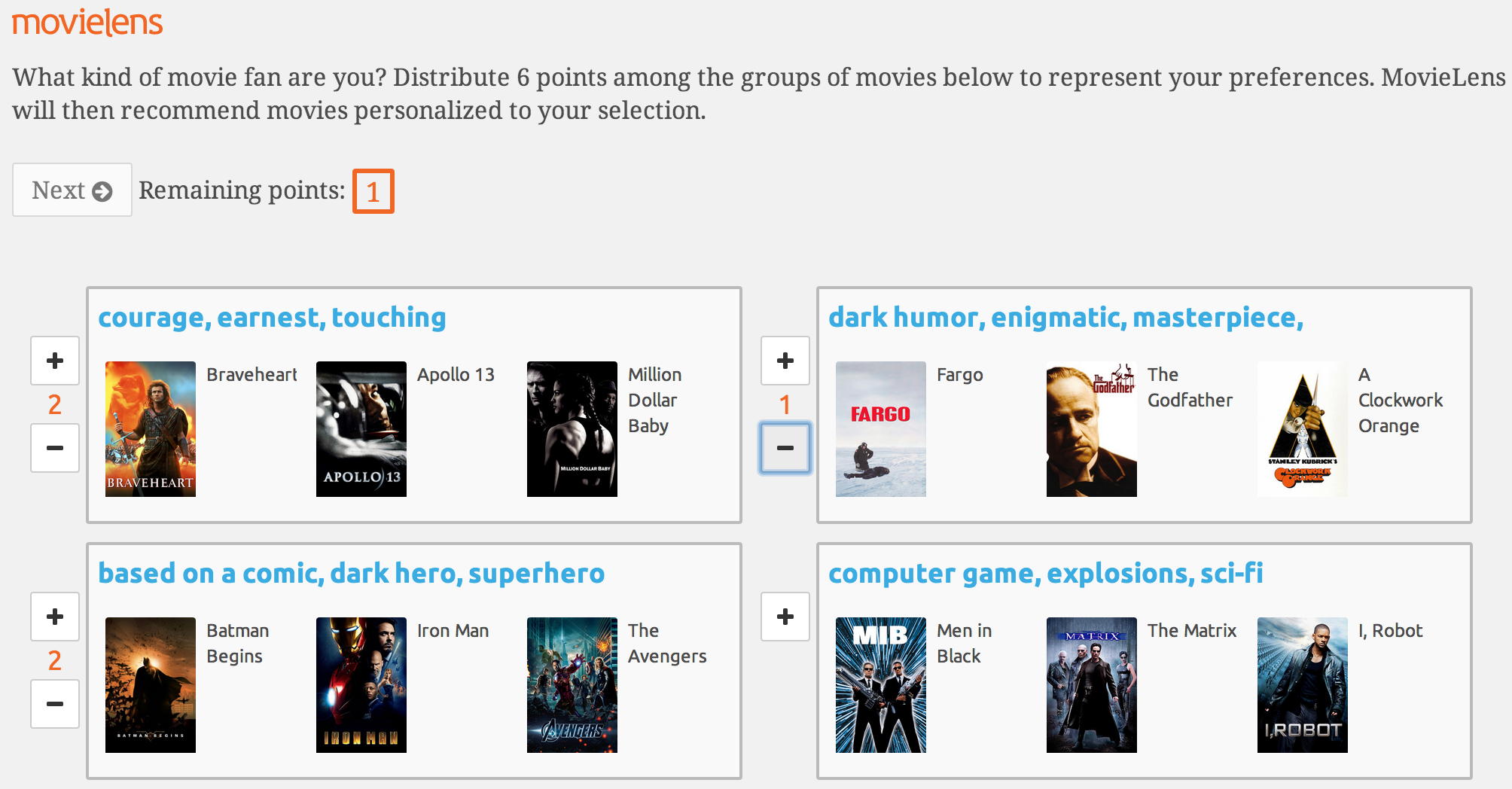}
    \caption{Web interface for group-based preference elicitation, \ie stage $R_0$. Users assign points to different movie groups. There is no individual movie ratings at this stage.}
    \label{fig:coldstart15}    
    \end{subfigure}
\begin{subfigure}{0.85\textwidth}
\centering
    \includegraphics[width = 0.75\linewidth]{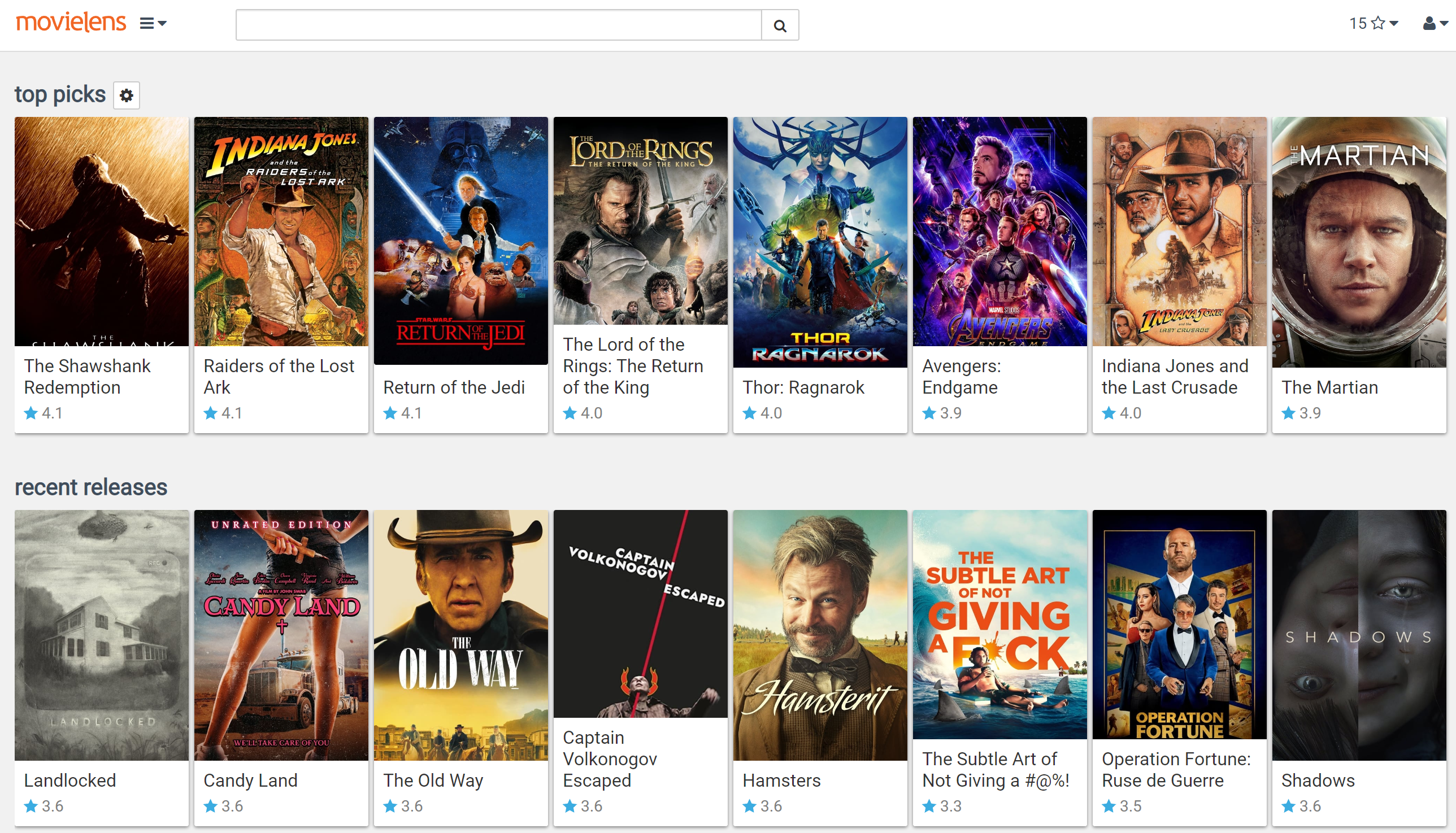}
    \caption{Web interface for item-based preference elicitation. Users assign ratings to the recommended movies in stages $R_1$ and $R_2$. The two stages are distinguished by the different internal recommendation algorithms used on MovieLens.}
    \label{fig:homepage}
\end{subfigure}
\caption{Screencaptures on MovieLens website for the three stages: (a) for stage $R_0$, (b) for stages $R_1$ and $R_2$. }
\label{fig:screencapture}
\end{figure*}

\subsubsection{Group-based preference elicitation} In the initial phase ($R_0$), when a new user registers on the platform, he/she is prompted to express his/her preferences on different movie groups by assigning points to each group (see Figure~\ref{fig:coldstart15}). The assigned points help to establish the initial pseudo-rating profile of a user \eg a user likes action movies more than documentary movies. Utilizing the pseudo-rating profile, the MovieLens main page displays a list of personalized recommendations from \textit{a restricted pool of representative movies}, denoted by $M_1$, mostly well-known movies. 

\subsubsection{Item-based preference elicitation} In the second stage ($R_1$), the platform seeks to obtain more fine-grained user preferences by prompting users to rate candidate movies from the initial personalized recommendation list (see Figure~\ref{fig:homepage}).
MovieLens requires users to surpass a certain number of ratings (\ie 15 movies) before they can access the full-scaled system ($R_2$). 

\subsubsection{Recommendation with preferred algorithm} At stage $R_2$, MovieLens allows users to choose their preferred recommendation algorithm; consequently users can continue to rate movies among the recommended titles in an iterative manner \eg submitting ratings of candidate movies shown on the current web page, and receiving the next list of candidate movies to rate on the next web page. The movies recommended on these pages are from $M_2$, a much larger pool of movies than $M_1$. 

The platform offers four distinct algorithms: (1) a non-personalized algorithm to recommend movies with the highest average ratings; (2) the same Pick-Groups recommender in $R_1$; (3) an Item-Item recommender to use cosine similarity on item-mean-centered rating vectors for recommendations; (4) an SVD recommender which utilize the FunkSVD algorithm~\cite{paterek2007improving} to capture latent patterns among the collected user-item interaction data. If a user does not choose any preferred recommendation algorithm, the default choice is Algorithm (3), \ie an item-item recommender. 

Note that, prior to Version 4, MovieLens recommended movies for users to rate strictly based on user-personalized predicted values, by its underlying collaborative filtering algorithms. With Version 4, the platform incorporates a popularity factor alongside the predicted values, in the final recommendations.
\begin{equation}
    \hat{s_{ui}}= 0.9 \cdot rank(s_{ui}) +0.1 \cdot rank(p_i)
\end{equation}
Specifically, the final rank score $\hat{s_{ui}}$ of a movie is a linear combination of its predicted rank score $s_{ui}$ by a recommender and its popularity score $p_{i}$, where $p_{i}$ is derived based on the number of ratings a movie received in the past year. Here, the $rank()$ function normalizes its input to $[0, 1]$ with 1 being highest rank.

We note that our review of the user interaction (or web interfaces) with the MovieLens platform is mainly on movie ratings. The MovieLens platform provides many other features like search for movies and add new movies~\cite{harper2015movielens}. As our key focus is the data collection of user-movie ratings, we do not cover these additional features offered by the MovieLens platform.

\subsection{The \dataset Dataset} 
\label{ssec:dataset1518}

As aforementioned, the MovieLens platform went through different versions. Hence, the interaction mechanism between users and the platform changes accordingly, which may affect the resultant data collected. To minimize the potential impact of different versions, we extract user interaction data from the last four years  (2015-2018) in the MovieLens-25M dataset, then remove duplicates. Note that, MovieLens-25M was released in 2019 and is currently the most recent and largest dataset ever released by GroupLens at the time of writing.\footnote{\url{https://grouplens.org/datasets/movielens/}} The reason for selecting the interactions that happened from 2015 to 2018 is that the new UI design and new recommendation engine (\ie Version V4) were deployed on MovieLens in November 2014. Since then, the MovieLens platform's version consistency is maintained. As a result, from the released MovieLens-25M dataset, we curate a subset for this study and we name it the \dataset dataset, for having user-item interactions that were recorded during the four years from 2015 to 2018. The \dataset dataset contains about 4.2 million user-item interactions (see Table~\ref{tab:Dataset}).

In the curated \dataset dataset, we only retain ratings from the users whose entire rating history falls within the 4-year period. This is to guarantee a comprehensive view of every user in this newly curated dataset. In addition, we remove the less active users who have fewer than 35 interactions with MovieLens, from the collected data. By setting 35 interactions as a threshold, we have good information about users' behaviors in the $R_1$ stage (\ie 15 movie ratings) and also the subsequent stages (\ie at least another 20 movie ratings). Note that, in the original MovieLens-25M dataset, each user has at least 20 ratings. In our analysis, we consider the first 15 ratings to be different from the rest, because the first 15 ratings obtained in stage $R_1$ are from a smaller pool of candidate movies (\ie $M_1$ in Figure~\ref{fig:process}).

\begin{table}[t]
  \center
   \caption{\dataset is a curated dataset from the MovieLens-25M dataset, where a user's all interactions were recorded from 2015 to 2018, and each user has at least 35 ratings. }
  \label{tab:Dataset}
  \renewcommand\arraystretch{1.2}
  \small
  \begin{tabular}{l|c|c|c|c|c}
      \toprule
      Dataset & \#Users &  \#Items &  Avg \#Ratings per user & Avg \#Ratings per item & \#Interactions\\
      \midrule
      \dataset & 24,812 & 36,378 & 170.3 & 116.2 & 4.2M\\
      \bottomrule
  \end{tabular}
 
\end{table}

\paragraph{The four yearly subsets.} In a typical RecSys evaluation, a dataset is used as a whole without considering when the user-item interactions occur; in other words, the global timeline of user-item interactions is not considered~\cite{ji2023critical,sun23newlook}. As the MovieLens internal recommendation considers the popularity score of movies in the past year, we follow the same time scale and set the evaluation time window as one year and conduct independent comparative experiments year by year, from 2015 to 2018.  That is,  we divide the \dataset into 4 subsets according to the global timeline, each subset for ratings made in one year, from 2015 to 2018. The data partition is based on a user's last rating. That is, if a user's last rating falls in Year 2016, then all of this user's ratings belong to the 2016 subset, regardless of whether some of the earlier ratings are made in 2016 or earlier. As most users complete all their ratings in a few days, most ratings are indeed made in the indicated year. For the completeness of the experiments, we also report the results obtained on the entire 4-year data when necessary.  

\section{Analysis on the \dataset Dataset}
\label{sec:datasetAnalysis}
We start with the analysis on the \dataset dataset from user perspective, particularly on the amount of time users spend on the platform. 
\begin{observation}
   On the MovieLens platform, users typically complete all their movie ratings within a very short time frame, ranging from a single day to a few days. 
\end{observation}
We observe that 49.19\%, or nearly half of all users,  complete all their ratings within a single day. More than 85.6\% of all users complete all their ratings within 5 days. And in many cases, a few ratings from the same user are recorded within a few seconds. In some extreme cases, the ratings are recorded with the same timestamp, making it difficult to determine the order of interactions. Specifically, in \dataset, 7.53\% of interactions share a same timestamp as another interaction in the same user rating sequence. 

Note that, this observation is not new. Similar observations on MovieLens dataset temporal patterns are also made by \citet{woolridge2021sequence}. In fact, the creators of the dataset also expressed concerns about the value of timestamps in their 2015 paper~\cite{harper2015movielens}. While this observation is not new, it is worth mentioning here that MovieLens only records user ratings of movies, but not when or where the user interacts with a particular movie. In other words, the rating of a movie is based on a user's memory after he/she has watched the movie before interacting with MovieLens. 

\subsection{Item-based Receptive Field Expansion}

\begin{table}[t]
    \centering
    \caption{Distribution of users' receptive fields at different stages. The receptive field is the number of candidate movies derived from all users during a specific stage (\eg the 16th to the 30th ratings indicated by 16-30) in the yearly dataset. The number of candidate movies derived from all ratings are listed in the last column.}

    \begin{tabular}{c|c|cccccc|c}
        \toprule
         \multirow{2}{*}{Year} & \multirow{2}{*}{\#User}&\multicolumn{7}{c}{Receptive field (\#Movies) at different stages} \\ \cmidrule{3-9}
         & & 1-15&16-30&31-45&46-60&61-75&76-90& all ratings\\
         \midrule
          2015	&2,544 &	  2,240	& 2,919      &	  3,470	&	3,851 &4,136	&4,468 &15,160	\\ 
           2016	&3,619 & 2,799	& 3,385      &	  3,898	&	4,173 &4,503	&4,811 &20,470	\\ 
           2017	&3,529 & 3,312	& 3,876      &	  4,257	&	4,415 &4,694	&4,919 &22,899	\\ 
           2018	&3,599 & 3,690	& 4,088      &	  4,491	&	4,692 &4,937	&5,250 & 27,596	\\ 
         \bottomrule
    \end{tabular}
    \label{tab:moviefield}
\end{table}

\begin{table}[t]
    \centering
    \caption{IoU of users' receptive fields between consecutive stages. Here "15 vs 30" refers to the IoU between the two stages "1-15" vs "16-30"; the same naming conversion applies to the rest columns.  }

    \begin{tabular}{c|ccccc}
        \toprule
         \multirow{2}{*}{Year} &\multicolumn{5}{c}{Intersection over Union } \\ \cmidrule{2-6}
          & 15 vs 30&30 vs 45&45 vs 60&60 vs 75&75 vs 90\\
         \midrule
          2015	 &	  0.4050 	& 0.4735       &	  0.4950 	&	0.5251  &0.5269 		\\ 
           2016	 & 0.4035 	& 0.4610       &	 0.5016 	&	0.5112  &0.5211 		\\ 
           2017	 & 0.3866	& 0.4456       &	 0.4693 	&	0.4848  &0.4936	\\ 
           2018	 & 0.3850	& 0.4296       &	 0.4653 	&	0.4821 &0.4854 		\\ 
         \bottomrule
    \end{tabular}
    \label{tab:ioufield}
\end{table}

From the data collection mechanism shown in Figure~\ref{fig:process}, for a given user, the pool of candidate movies that a user can rate changes at different stages.  We borrow the concept of "receptive field" to understand the MovieLens dataset. In Convolutional Neural Networks (CNNs), the receptive field of a neuron refers to the size or extent of the region in the input that affects the output of that neuron. Essentially, it's the area of the image that a neuron "sees" or processes. In our context, the receptive field is essentially the union set of items that users have interacted with at different stages of the process, acting as a proxy of the pool of recommended items.\footnote{The optimal method for deriving the receptive field in this context is by utilizing movies listed in the impressions. However, since the MovieLens dataset lacks impressions, the rated movies are employed as a proxy.}

To facilitate meaningful comparisons across different user interaction stages, it's important to maintain a constant number of interactions at each stage from a fixed group of users, analogous to one layer in CNN. We fix the number of interactions to 15 at each stage. The median number of interactions per user is 98 in our dataset. Hence, we use 90 (the closest multiple of 15) as a threshold for user filtering, to ensure that we have a good number of users to be included in this study. Accordingly, users who have at least 90 interactions are used in this study. Then we derive the receptive field and compute their similarities for this fixed group of users, at different stages for \textit{each batch of 15 ratings as one stage}.

Table~\ref{tab:moviefield} reports users' receptive fields in a matrix format. The reported "receptive field" is the number of unique movies that are ever rated by all users during that stage (\eg from the 16th rating to the 30th rating for stage 16-30).  The number of unique movies across all rating stages are listed in the last column of the table, for reference.   

\begin{observation}
Along the interaction progresses, from the initial stage (1-15) to the last stage (76-90), users are provided with a broader range of movies to rate, leading to continually expanding their receptive field. 
\end{observation}
This observation holds on all four years. Further, for a given stage \eg 1-15, the receptive field increases over time from 2015 to 2018, as more movies are released over time. We further use Intersection over Union (IoU) as the similarity metric to measure the overlap between two receptive fields, reported in Table~\ref{tab:ioufield}. 
\begin{observation}
There is an increasing overlap of receptive fields across various stages of user interactions, ranging from the initial stage (1-15) to the last stage (76-90). This indicates that the divergence in user preferences is more pronounced during the initial and intermediate stages of interactions.
\end{observation} 
The later stages of user interactions are gradually approaching the threshold of the system's candidate items.  From 2015 to 2018, the receptive field increases for any given particular stage from (1-15) to (76-90). Accordingly, the IoU of receptive fields at consecutive stages drops slightly along the year dimension.

\subsection{Group-based Preference Invariance}

We now focus on the possible user preference changes during the interaction.

\begin{observation}
While the user's receptive field expands during the interaction process, their group-based preferences remain relatively fixed.
\end{observation}

Recall that in $R_0$, users allocate points to various movie groups. Based on these allocations, users complete the first 15 ratings in $R_1$. Accordingly, the genres of these 15 movies are expected to be highly focused, \eg similar to the groups of movies assigned with higher points. Among the first 15 ratings of each user, we derive the top 3 genres that the user is most interested in. The result shows that nearly 90\% of the first 15 rated movies belong to 3 genres they are most interested in, across the four years from 2015 to 2018 (see Table~\ref{tab:PrefInvariance}). 

\begin{table}[t]
  \center
  \caption{The ratio of movies falling within the top 3 genres of interest, among the first 15 ratings and the last 15 ratings, respectively.}
  \label{tab:PrefInvariance}
  \begin{tabular}{c|cc}
      \toprule
  Year  & First 15 ratings & Last 15 ratings \\
      \midrule
      2015 & 0.8961 &0.7464 \\
      2016 & 0.9072 &0.7435 \\
      2017 & 0.8957 &0.7322 \\
      2018 & 0.8912 &0.7253 \\
      \bottomrule 
  \end{tabular}
\end{table}

Further, we calculate the ratio of each user's last 15 ratings containing movies from the same three genres, to investigate potential changes in preferences after numerous additional ratings in between. Note that, the pool of movies that appear in the last 15 ratings is nearly double the size of that in the first 15 ratings. Reported in Table~\ref{tab:PrefInvariance}, more than 70\% of the last 15 rated movies fall in the top 3 genres that users are initially interested in.\footnote{We have also conducted experiments to study the impact of the number of ratings a user has, to the ratio of the last 15 movies belonging to the top 3 genres. The results show that the ratios range from 67.4\% to 70.7\% in different years, for the top 25\% users having the most number of ratings. These numbers are not far away from the 70\% reported in Table~\ref{tab:PrefInvariance}. In other words, the number of ratings a user has does not impact user conformity to the top interested genres.}

\section{Experiments}
\label{sec:expSetting}

In this section, we design experiments to investigate the impact of the interaction collection process to recommender models. We first present the baseline models and evaluation metrics used in our empirical analysis. Note that, as the train/test instances change as required by the experimental setting, we  detail the train/test split under each set of experiments. Codes used in our experiments are  available online.\footnote{ \url{https://github.com/FrankYuchen/RevistMovieLens}}

\subsection{Baseline and Evaluation Metric}
We select seven widely used baselines from four categories: (1) memory-based methods \ie MostPop and ItemKNN~\cite{DBLP:journals/tois/DeshpandeK04}; (2) latent factor method PureSVD~\cite{DBLP:conf/recsys/CremonesiKT10}; (3) non-sampling deep learning method  Multi-VAE~\cite{liang2018variational}; and (4) sequence-aware deep learning methods SASRec~\cite{kang2018self},  TiSASRec~\cite{li2020time}, and  Caser~\cite{tang2018personalized}. 

More specifically, \textbf{MostPop} is a non-personalized method that recommends items based solely on their popularity. In our experiments, we follow the widely adapted popularity definition, to indicate an item's popularity by the number of interactions it receives in training set, \ie within one year in the four yearly datasets, and within four years for the entire dataset. \textbf{ItemKNN} is a collaborative filtering algorithm that focuses on finding similar items based on users' past interactions. It is an extension of the K-Nearest Neighbors (KNN) algorithm, specifically tailored for recommendation systems. \textbf{PureSVD} is a matrix factorization technique used for recommendation systems,  which decomposes a user-item interaction matrix into three smaller matrices: user factors, singular values, and item factors. \textbf{Multi-VAE} is a deep learning-based recommendation algorithm that leverages the power of variational autoencoders (VAEs) to learn latent representations of users and items.

For sequence-aware models, \textbf{SASRec},\textbf{TiSASRec}, and \textbf{Caser} are chosen as the representative models to model meaningful sequential patterns. In particular, SASRec uses a self-attention based network to model long term dependencies within user-item interaction sequence. TiSASRec further incorporates time interval information between user behaviours in sequence modelling process, building on top of the self-attention network. Caser, on the other hand, employs a convolutional neural network architecture (CNN) to capture sequential patterns.

As for the implementation and parameter setting, non-sequential recommendation algorithms are supported by the open-source recommendation library DaisyRec 2.0~\cite{sun2022daisyrec}. This is to ensure the reproduction of all running details and studied settings. We have verified the implementations on the MovieLens-1M dataset, and obtained consistent results  using the configuration provided in the DaisyRec paper~\cite{sun2022daisyrec}.

Regarding SASRec and TiSASRec, we employ third-party implementations,\footnote{\url{https://github.com/pmixer/SASRec.pytorch}, \url{https://github.com/pmixer/TiSASRec.pytorch}} endorsed by the original authors.\footnote{The authors' GitHub pages provide links to the PyTorch implementations.}  For Caser, we use the PyTorch version provided by the authors.\footnote{\url{https://github.com/graytowne/caser_pytorch}} 
Third-party implementations, while potentially boosting algorithm popularity, may exhibit
non-negligible deviations~\cite{thirdPartyImplementation} from the original author’s official implementation due to nuanced adjustments in the coding. To ensure the implementations are not flawed, we provide evaluation results on the standard MovieLens-1M dataset in Table~\ref{tab:evaML1M} in Appendix~\ref{sec:appdx}. The results suggest that our implementations produce comparable results as those reported in their original papers.

To search for optimal hyper-parameters for all recommendation algorithms, Bayesian HyperOpt is employed to optimize the hyper-parameters concerning NDCG@10 for each model on the dataset over the course of 30 trials~\cite{ferrari2019we}. 

\paragraph{Evaluation Metrics.} HR@$k$ and NDCG@$k$ are the two metrics used to evaluate the effectiveness of ranked results in our experiments. HR@k focuses solely on whether the relevant items present among the top $k$ recommended items, while NDCG@k takes into account both the presence and ranking of the relevant items within the top-$k$ results. As suggested by~\cite{krichene2022sampled}, we use the \textit{full-rank} version of both metrics. That is,  we rank all candidate items, instead of sampling a fixed number of negative items for ranking. The number of the top recommended items $k$ is set to 10. We mainly discuss results measured by HR@10 as similar observations hold for results by NDCG@10.

\subsection{Impact of Interaction Context at Different Stages}
\label{ssec:IntContext}

In view of the observations made from our earlier analysis, the first $15$ ratings of a user shall be treated differently from the subsequent interactions, when exploring recommendation models on MovieLens dataset. Accordingly, we conduct an ablation experiment with the removal of the first $15$ ratings of each user in the \dataset dataset. For comparison, we also report the results of removing randomly sampled $15$ ratings, and the removal of the last $15$ ratings from a user's training instances, respectively. For the experiments that randomly remove $15$ ratings, we repeat the experiments three times with different seeds and get the average recommendation performance to reduce random error. In this set of experiments, we  follow the leave-last-one-out data partition scheme. The removal of $15$ ratings only applies to the training set, for all three types of removal. We keep the validation set (the penultimate interaction) and the test set (the last interaction) unchanged for a fair comparison.\footnote{Note that, as every user has at least $35$ interactions in the \dataset dataset, removal of $15$ ratings means every user has at least $20$ ratings left.}

\begin{figure*}[t]
    \centering
    \begin{subfigure}[t]{0.4\columnwidth}
    \includegraphics[width=\columnwidth]{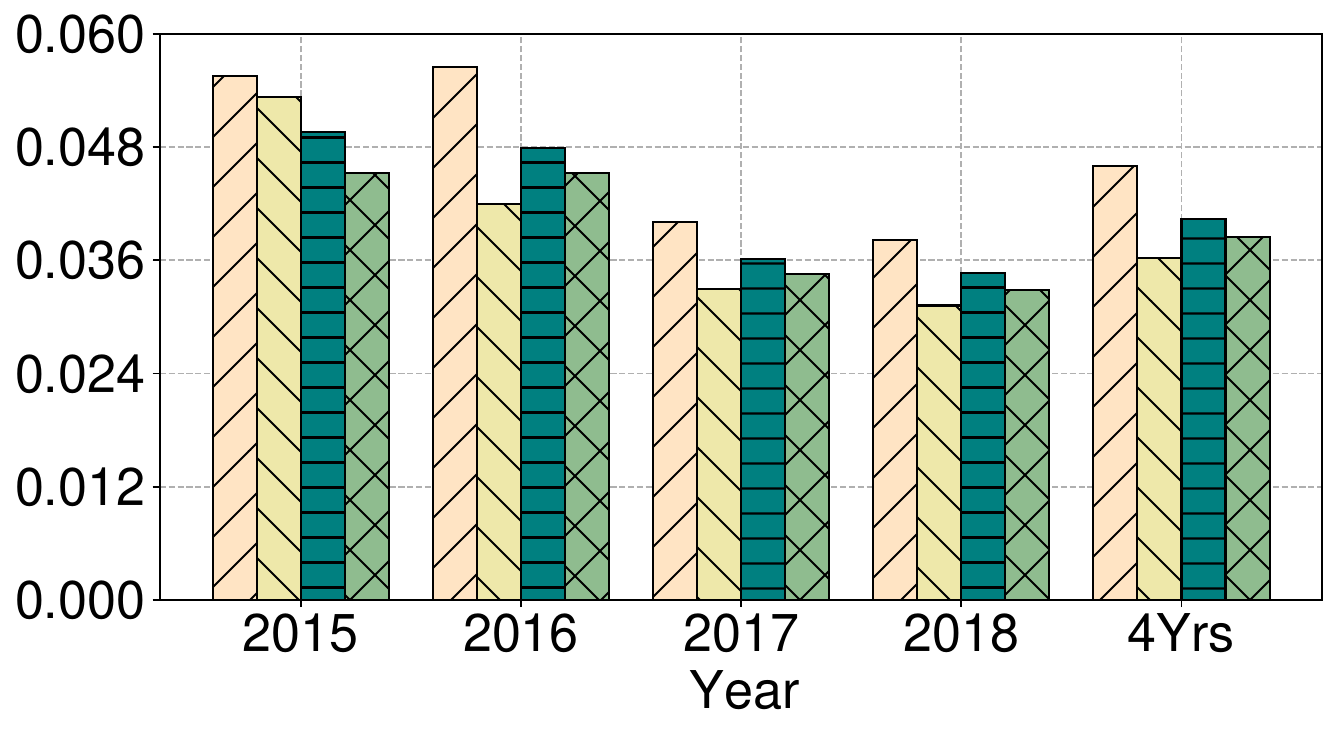}
    \caption{MostPop}
    \end{subfigure}
    \quad
    \begin{subfigure}[t]{0.4\columnwidth}
    \includegraphics[width=\columnwidth]{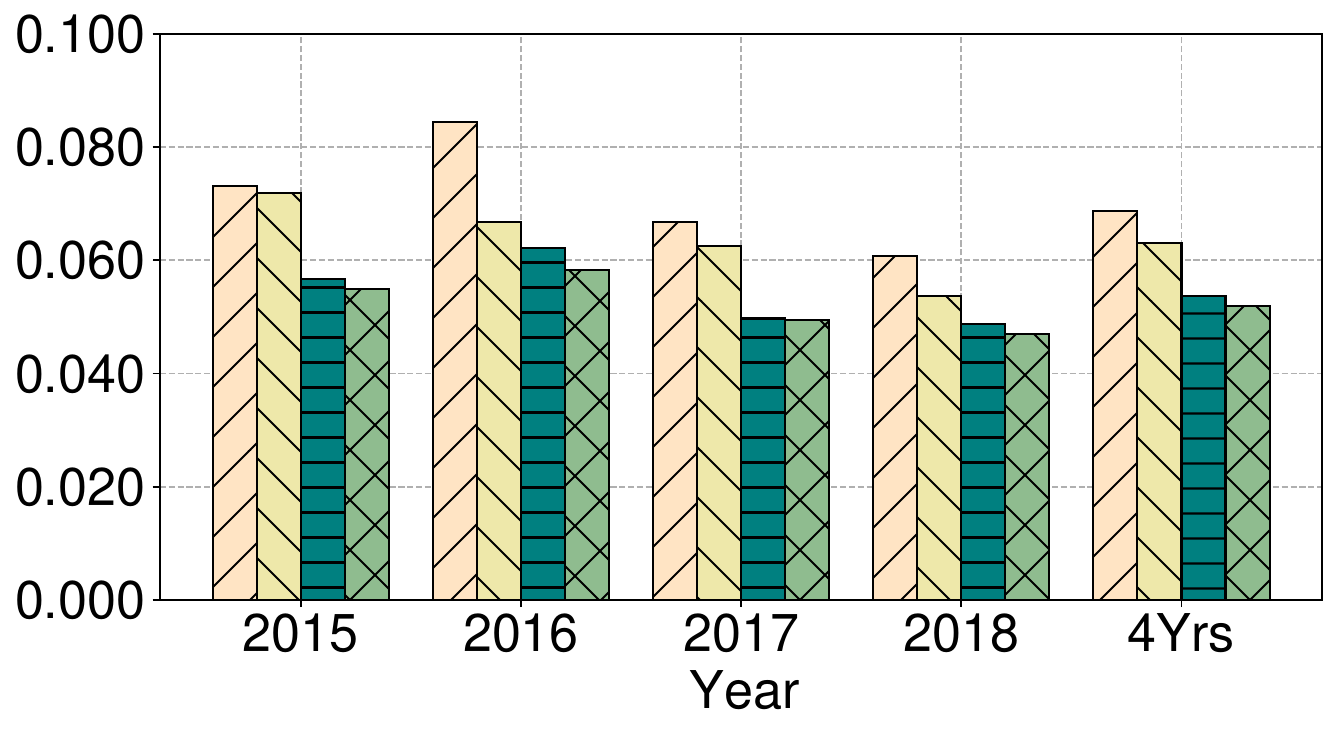}
    \caption{ItemKNN}
    \end{subfigure}
    \quad
    \begin{subfigure}[t]{0.4\columnwidth}
        \includegraphics[width=\columnwidth]{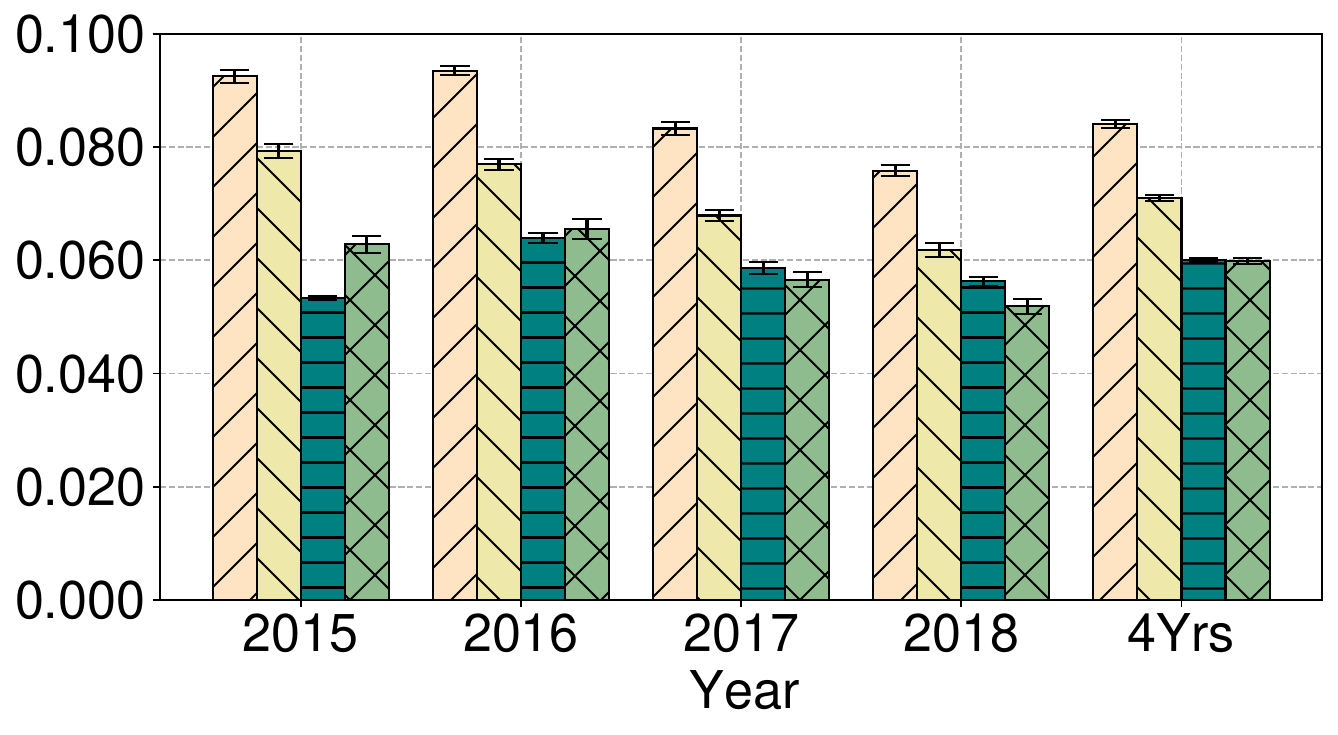}
        \caption{SVD}
    \end{subfigure}
    \quad
    \begin{subfigure}[t]{0.4\columnwidth}
        \includegraphics[width=\columnwidth]{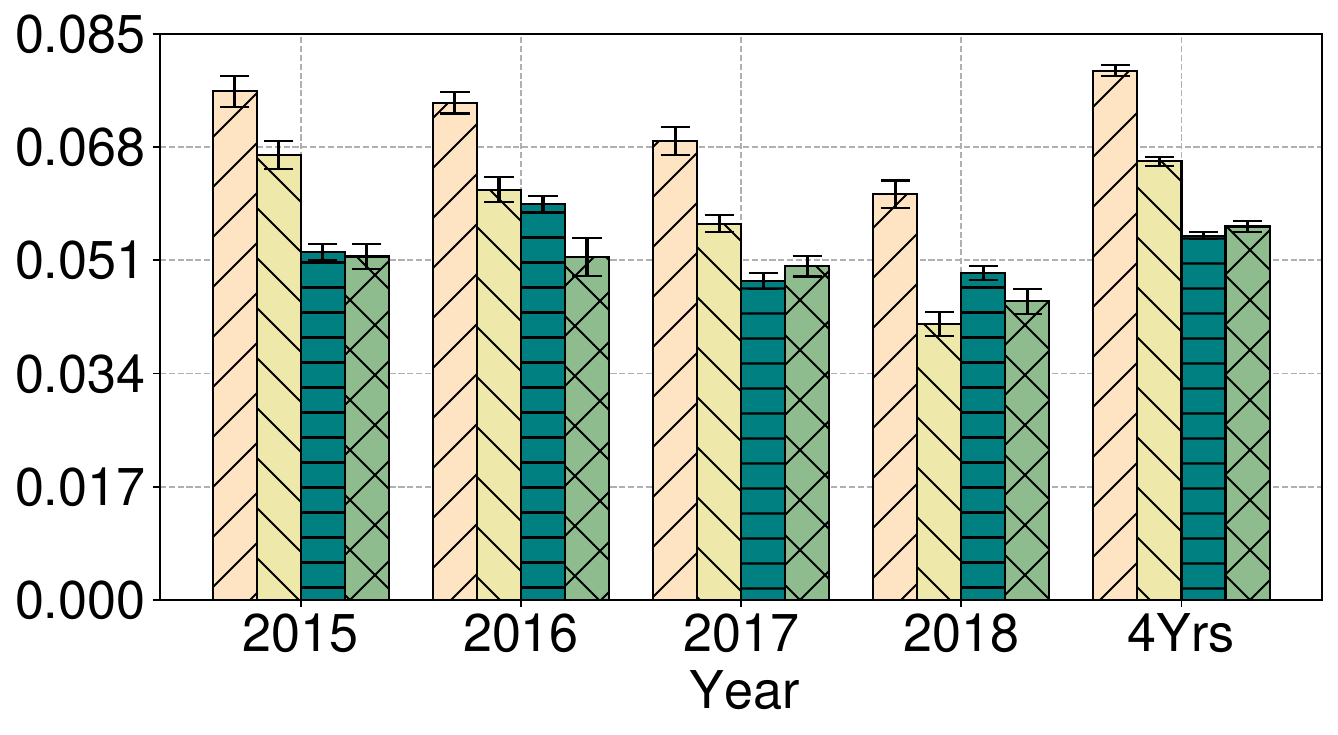}
        \caption{Multi-VAE}
    \end{subfigure}
    \quad
    \begin{subfigure}[t]{0.4\columnwidth}
        \includegraphics[width=\columnwidth]{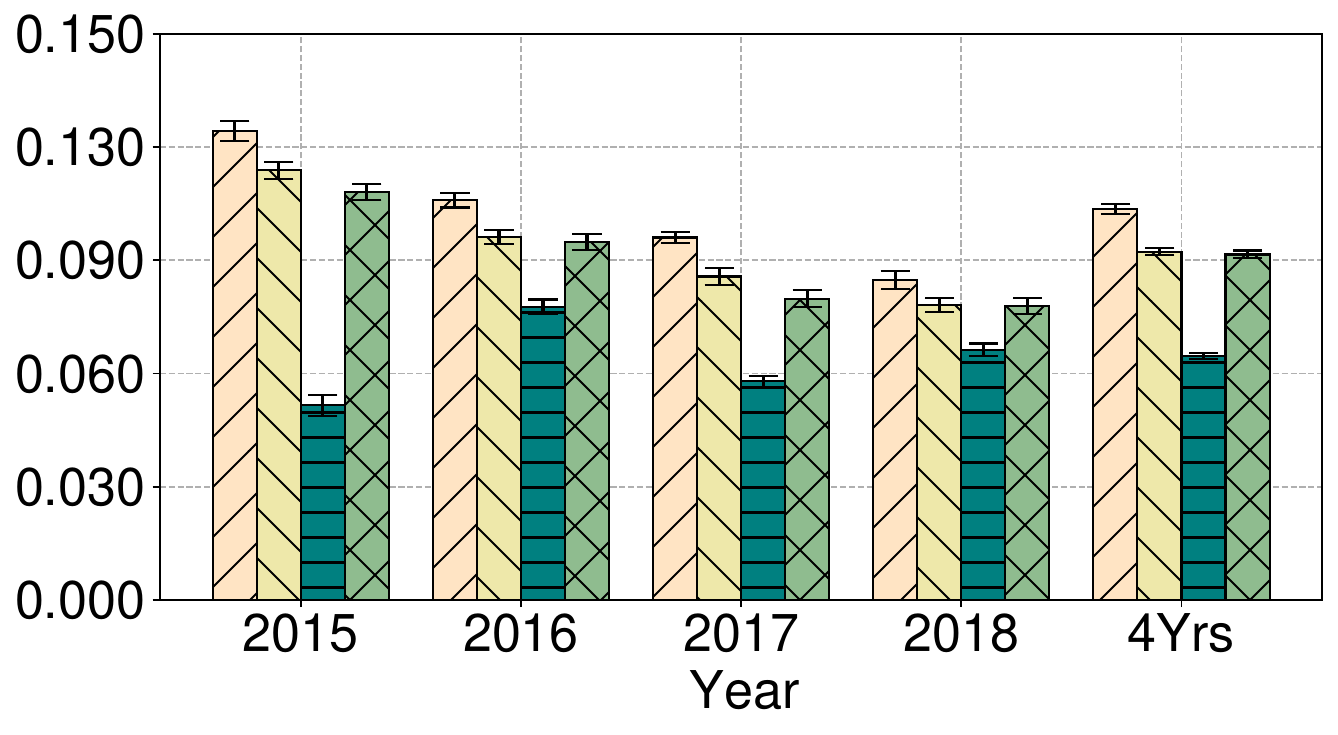}
        \caption{SASRec}
    \end{subfigure}
        \quad
    \begin{subfigure}[t]{0.4\columnwidth}
        \includegraphics[width=\columnwidth]{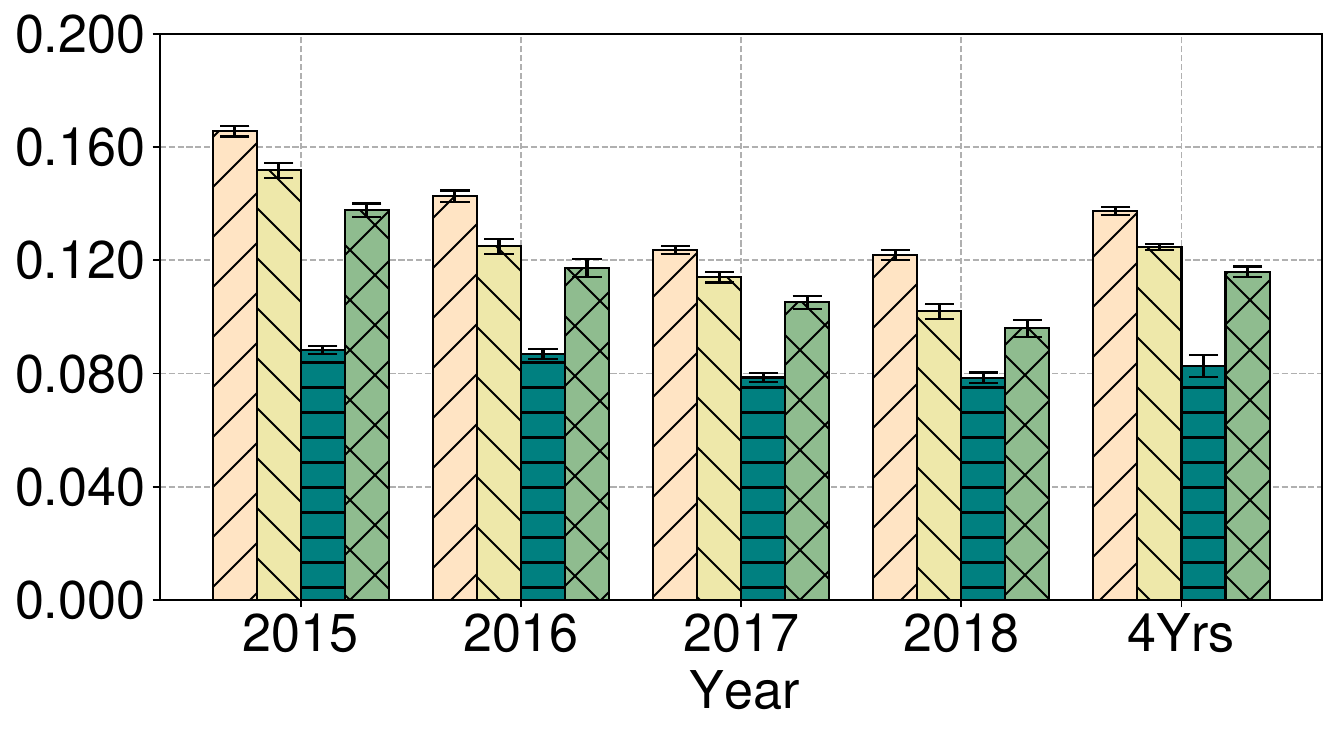}
        \caption{TiSASRec}
    \end{subfigure}
    \quad
    \begin{subfigure}[t]{0.4\columnwidth}
        \includegraphics[width=\columnwidth, trim = -0.8cm 0 0.8cm 0 ]{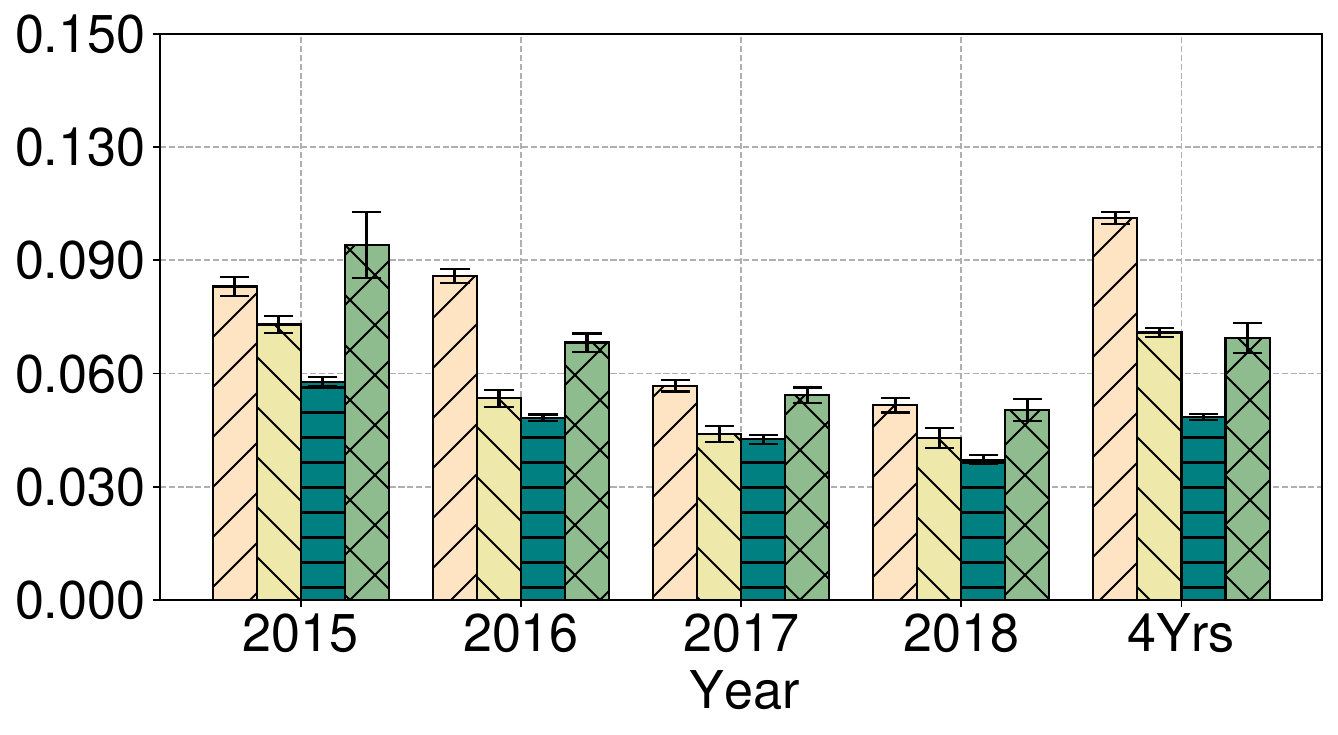}
        \caption{Caser}
    \end{subfigure}
    \quad
    \quad
    \begin{subfigure}[t]{0.4\columnwidth}
        \includegraphics[scale=0.2, width=\columnwidth, trim = -0.5cm -1.5cm 0cm 0cm]{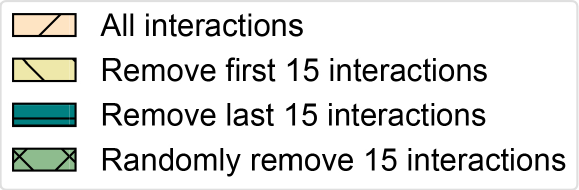}
    \end{subfigure}     
    \caption{The HR@10 results of removing 15 interactions from the training set of each user, on four yearly datasets, and the entire \dataset dataset covering all four years (indicated by "4 Yrs"). We evaluate three cases: (i) removal of the first 15 interactions, (ii) removal of the last 15 interactions, and (iii) removal of randomly sampled 15 interactions. The 95\% confidence intervals are indicated for the performance of non-deterministic algorithms.}
    \label{fig:allvsremove}
\end{figure*}

\begin{figure*}[t]
    \centering
    \begin{subfigure}[t]{0.4\columnwidth}
    \includegraphics[width=\columnwidth]{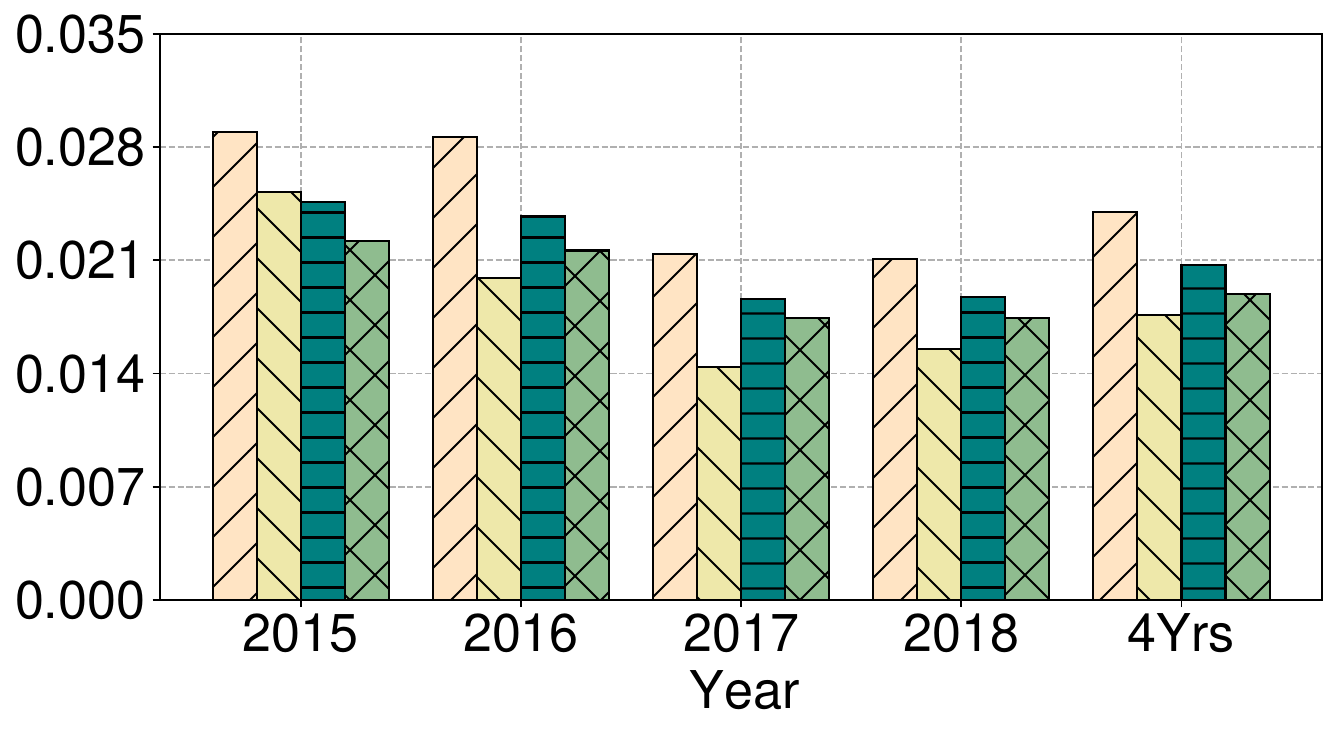}
    \caption{MostPop}
    \end{subfigure}
    \quad
    \begin{subfigure}[t]{0.4\columnwidth}
    \includegraphics[width=\columnwidth]{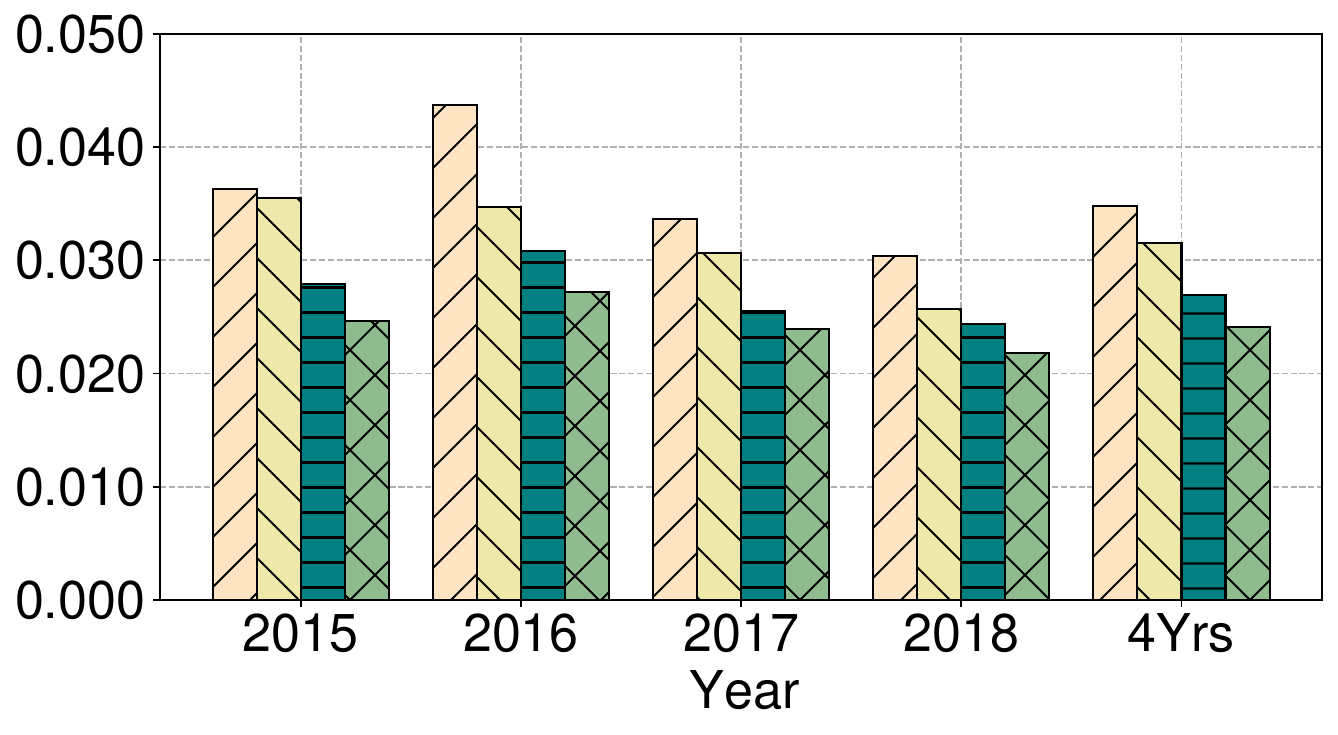}
    \caption{ItemKNN}
    \end{subfigure}
    \quad
    \begin{subfigure}[t]{0.4\columnwidth}
        \includegraphics[width=\columnwidth]{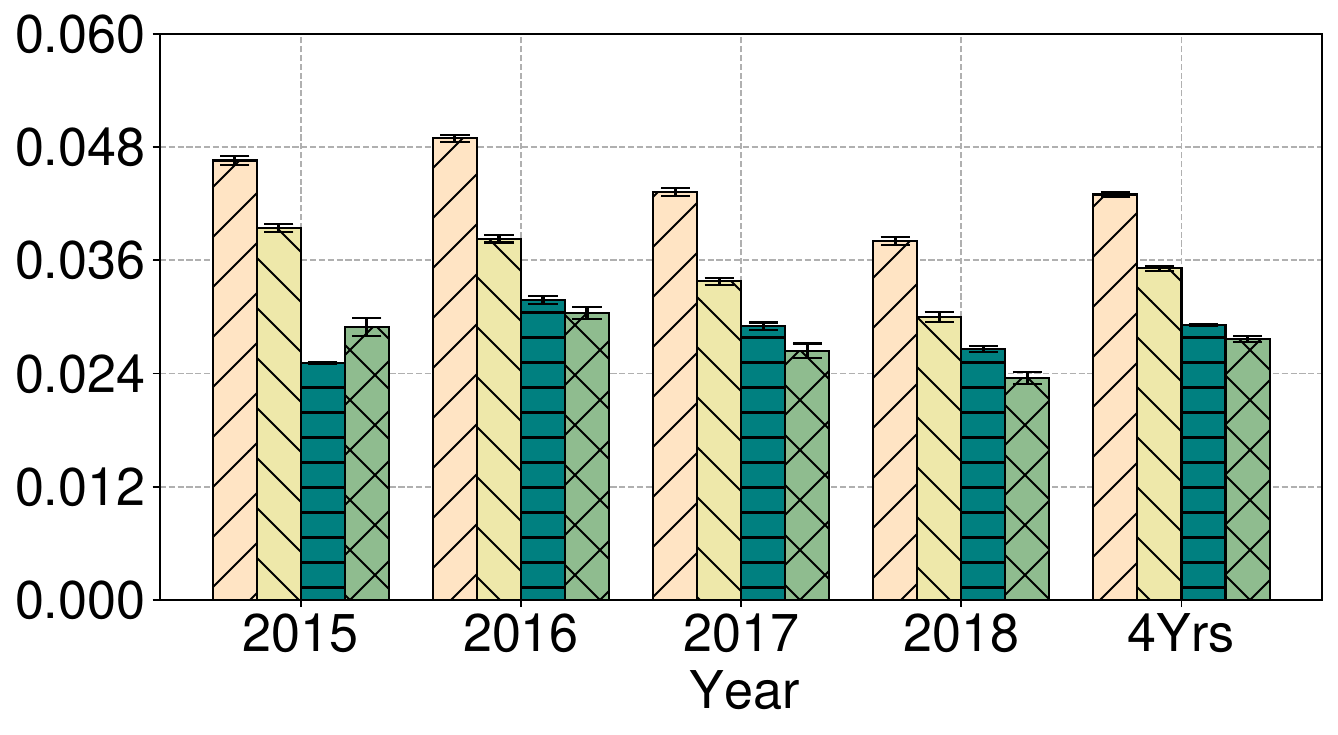}
        \caption{SVD}
    \end{subfigure}
    \quad
    \begin{subfigure}[t]{0.4\columnwidth}
        \includegraphics[width=\columnwidth]{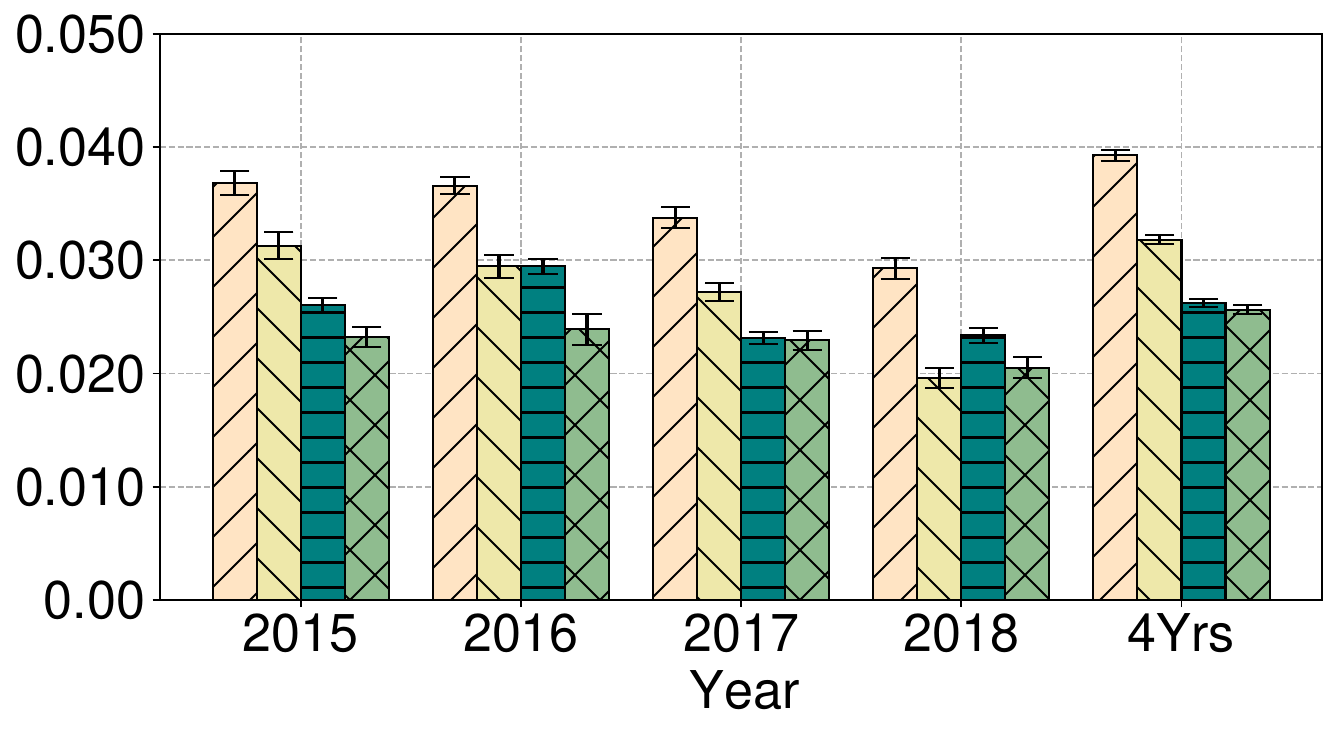}
        \caption{Multi-VAE}
    \end{subfigure}
    \quad
    \begin{subfigure}[t]{0.4\columnwidth}
        \includegraphics[width=\columnwidth]{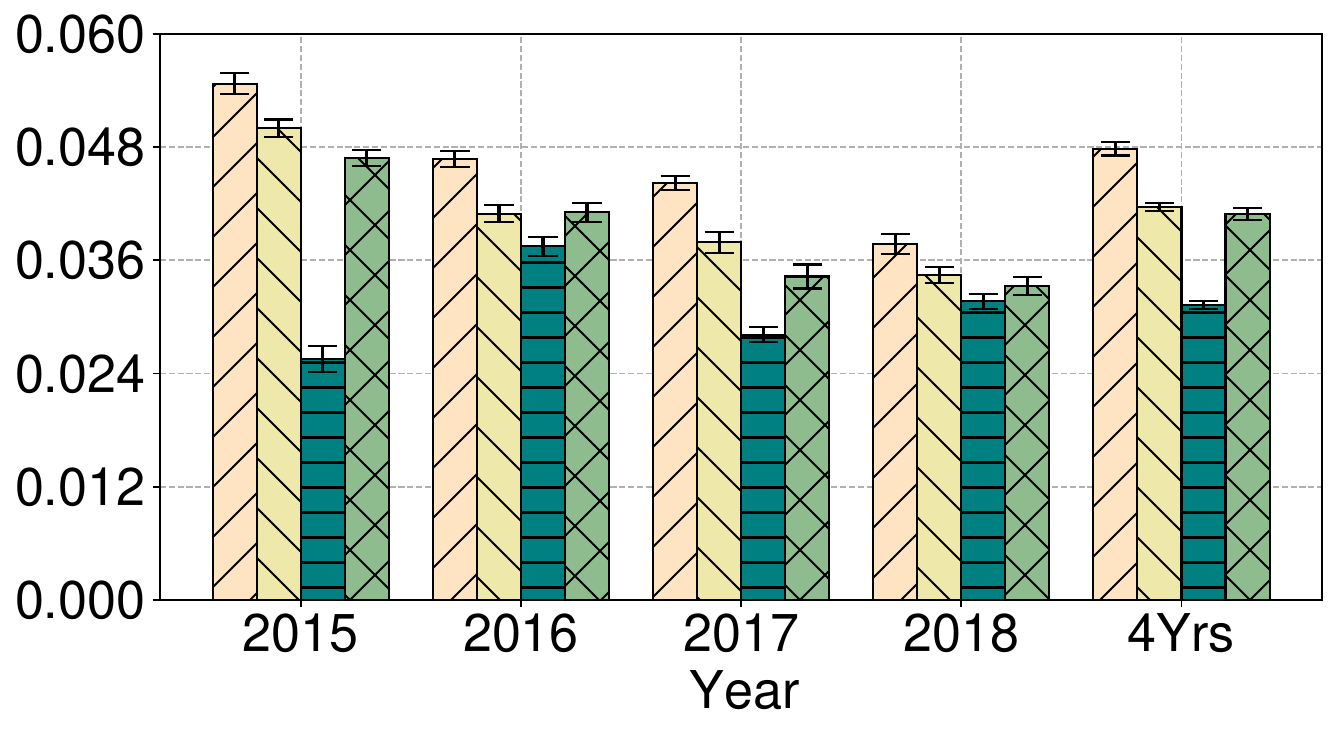}
        \caption{SASRec}
    \end{subfigure}
        \quad
    \begin{subfigure}[t]{0.4\columnwidth}
        \includegraphics[width=\columnwidth]{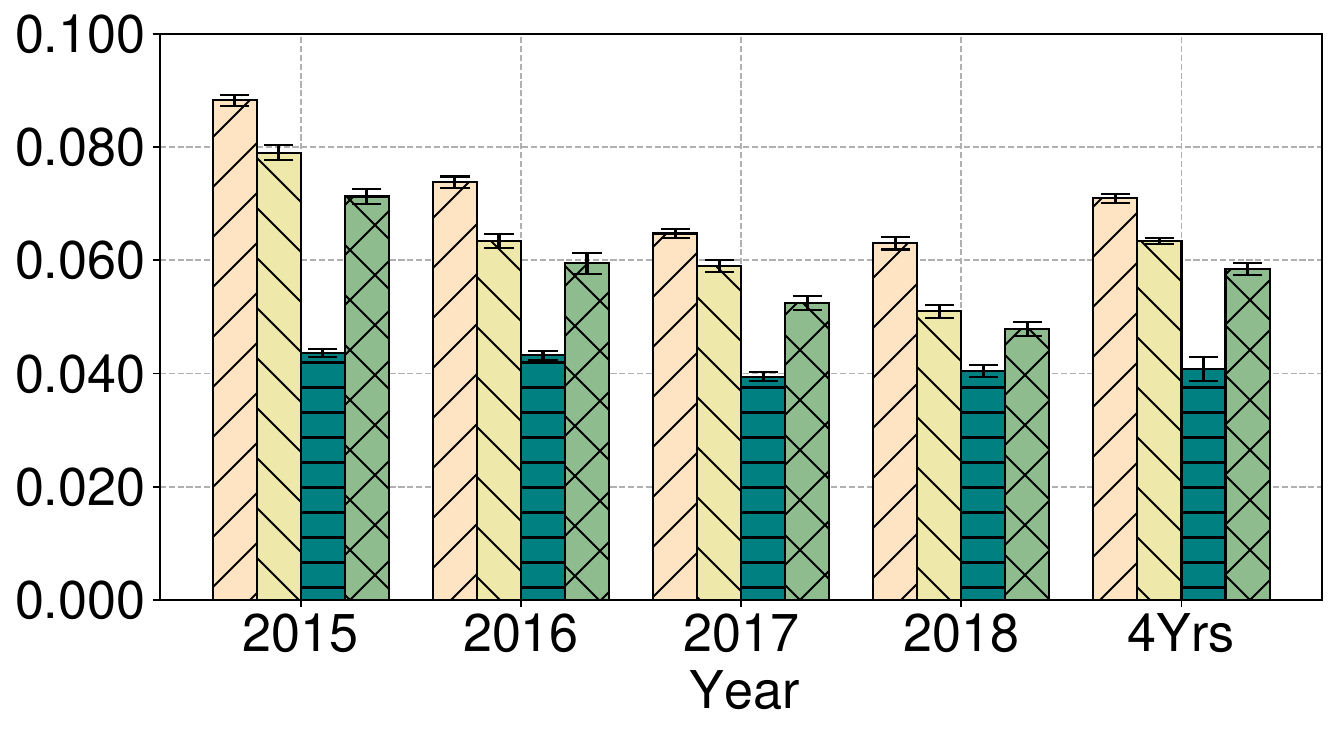}
        \caption{TiSASRec}
    \end{subfigure}
    \quad
    \begin{subfigure}[t]{0.4\columnwidth}
        \includegraphics[width=\columnwidth]{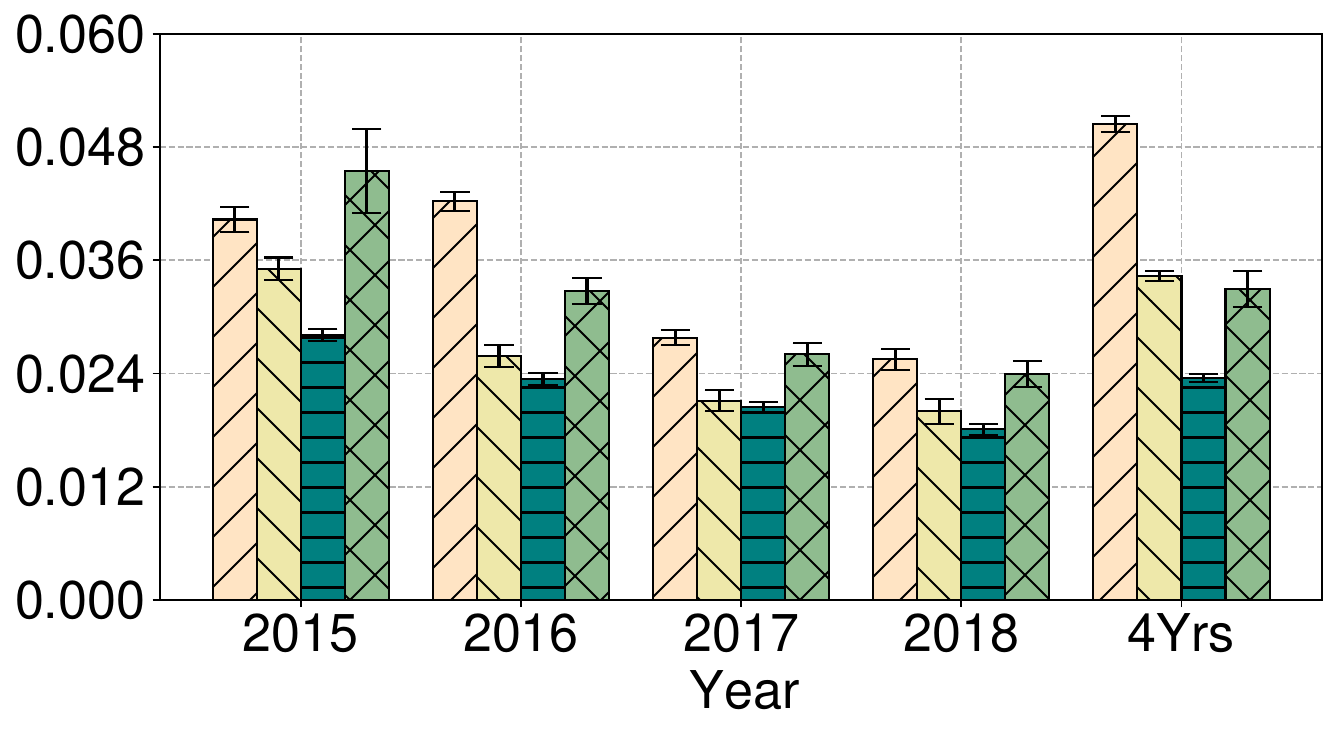}
        \caption{Caser}
    \end{subfigure}
    \quad
    \begin{subfigure}[t]{0.4\columnwidth}
        \includegraphics[scale=0.2, width=\columnwidth,trim = -0.5cm -1.5cm 0cm 0cm]{Figures/legend.pdf}
    \end{subfigure}    
    \caption{The NDCG@10 results of removing 15 interactions from the training set of each user, on four yearly datasets, and the entire \dataset dataset covering all four years (indicated by "4 Yrs"). We evaluate three cases: (i) removal of the first 15 interactions, (ii) removal of the last 15 interactions, and (iii) removal of randomly sampled 15 interactions. The 95\% confidence intervals are indicated for the performance of non-deterministic algorithms.}
    \label{fig:ndcg_allvsremove}
\end{figure*}

\begin{observation}
In general, recommendation accuracy drops when $15$ interactions are removed from the training set. The amount of drops in performance vary, depending on the types of interactions removed.
\end{observation}

According to the HR@10 scores plotted in Figure~\ref{fig:allvsremove}, the removal of $15$ interactions leads to a decrease in recommendation accuracy on all baselines. It is understandable, because partial information is lost due to the removal of training data.

In general, we observe that removing the last 15 interactions leads to poorer results compared to removing the first 15 interactions, particularly for the sequence-aware models. This is because the removal of the last 15 interactions results in the loss of the most recent information in the temporal context. It is thus more difficult to predict the most recent interaction, \ie the test instance in our leave-one-out setting. This observation applies for all baselines, except for MostPop. 

One reason for the opposite observation on MostPop is that the first 15 interacted movies are from a relatively small pool of movies pre-selected by the MovieLens platform~\cite{harper2015movielens}.
All users interact with movies in this small pool when they start to interact with MovieLens platform. Hence, the pool of movies is in an advantageous position of receiving more interactions from almost all users. MostPop, as a popularity-based method, is significantly impacted by the removal of the first 15 interacted movies, which are the popular movies.

Observe that, in Figure~\ref{fig:allvsremove}, we plot the results on the four yearly subsets, as well as the full dataset denoted by "4 Yrs". In general, on the yearly subsets, all baselines show a degradation trend in performance, as the candidate movie pool becomes larger from 2015 to 2018. For most baselines, except Multi-VAE and Caser, results on the full dataset are close to the average of the four yearly subsets. Multi-VAE and Caser achieve the best results on the full dataset. Similar observations hold on results measured by NDCG@10 plotted in Figure~\ref{fig:ndcg_allvsremove}.

\begin{figure}[t]
    \centering
    \includegraphics[width = 0.5\columnwidth]{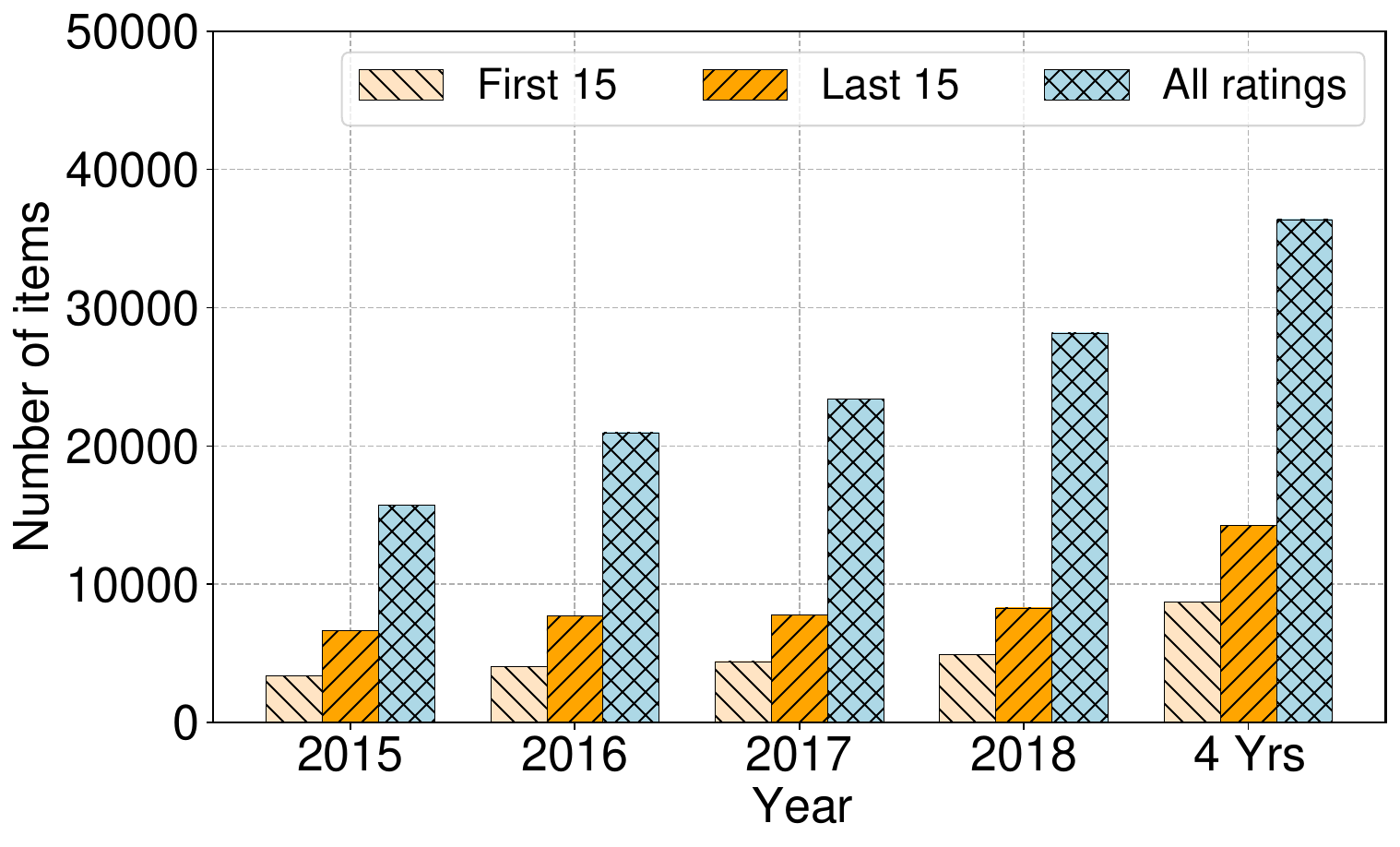}
    \caption{Comparison of the size of the candidate movie pool at different stages, on four yearly subsets, and also the entire \dataset dataset covering "4 Yrs".}
    \label{fig:itemnum}
\end{figure}

\begin{figure}[t]
    \centering
    \includegraphics[trim = 2cm 13cm 16cm 0cm, clip, width = 0.65\linewidth]{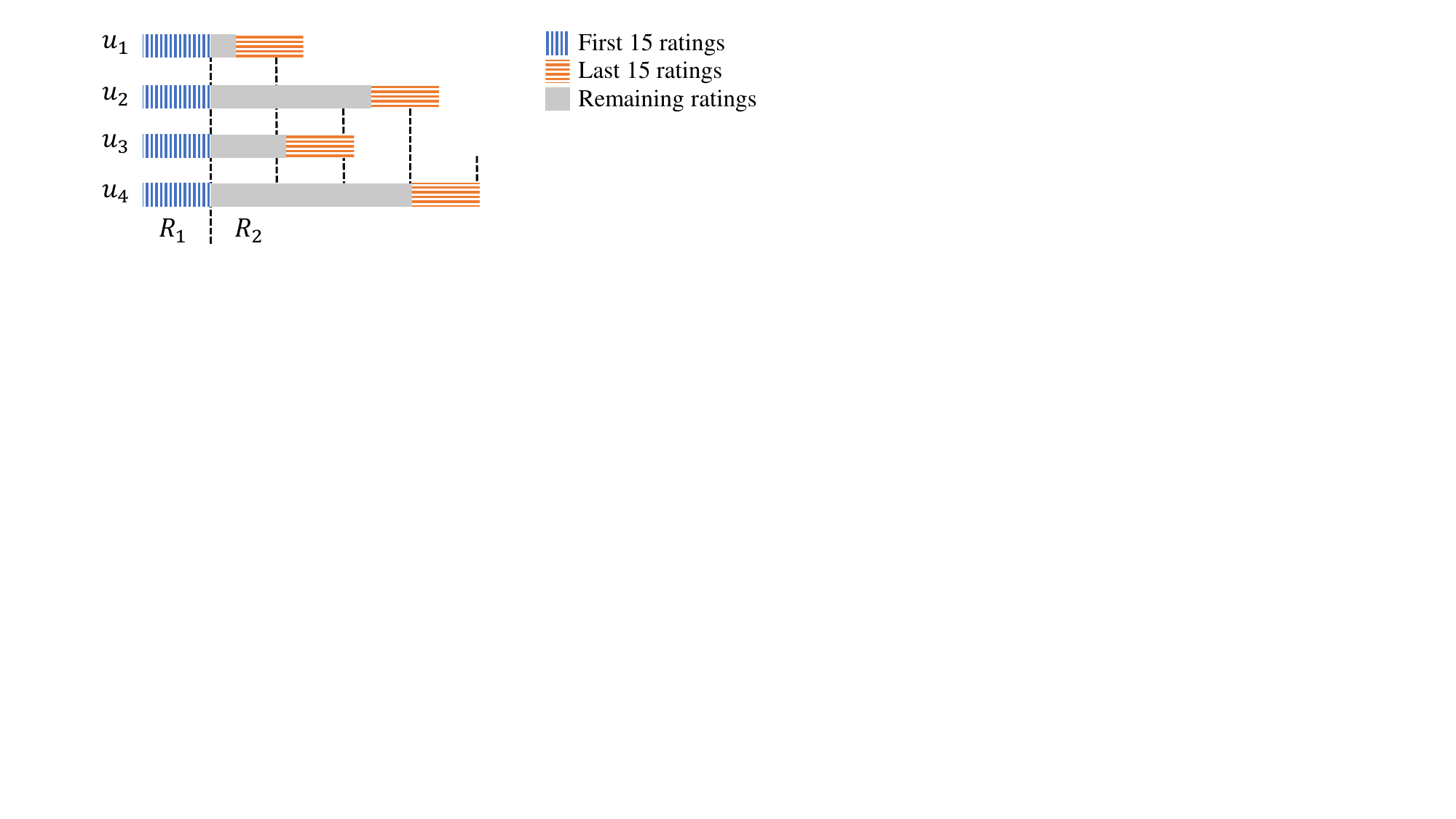}
    \caption{An example of four users with different numbers of movie ratings. The vertical dotted line is a simple indicator of user invoking recommendation algorithms during stage $R_2$ \ie the user submits her ratings on the current page and the system offers a new list of movies for rating on the next page. }
    \label{fig:userRating}
\end{figure}

In Figure~\ref{fig:itemnum}, we plot the number of unique movies that ever appear in a user's first 15 ratings, last 15 ratings, and all ratings, by year, in the \dataset dataset. The number of movies in each category is an indication of the "receptive field" in the recommendation.\footnote{Note that, the numbers reported in Figure~\ref{fig:itemnum} are different from that in Table~\ref{tab:moviefield}, because the latter was derived from a selected group of users (\ie those with at least 90 ratings).} The significant differences in size between the pools of movies for the first 15 ratings and the last 15 ratings are consistent with the previous observation of receptive field inflation. It also corroborates the MovieLens platform operator's description of the "Item-based Preference Elicitation" stage ($R_1$): The recommender for new MovieLens users, which uses movie group selection at $R_0$ to determine which movies to recommend, only recommends from a \textit{restricted pool} of movies.

Before we further analyze the impact of stage $R_1$ on recommendation system evaluation, it is necessary to revisit the problem of interaction context misalignment that existed in the previous ablation experiments. In Figure~\ref{fig:userRating}, we highlight the first and the last 15 movies (in different colors and patterns) rated by four example users. Note that, all the first 15 movies are rated in stage $R_1$, and the last 15 movies could be the results of the different number of runs of the selected recommendation algorithms. Each run of the algorithm will update what users see in the web page, and these will become candidate movies for user interaction and constitute the interaction context. Therefore, the interaction context of the test instance of each user is \textbf{different}. In fact, when a user has enough interactions with the platform, the user is able to explore more recommended candidate movies which are more personalized along his/her preference, as illustrated in the big tree on the left hand side of Figure~\ref{fig:process}, from a common root to more personalized preferences.

The relatively fixed pool of candidate items set by the MovieLens platform for users in the $R_1$ stage makes the first $15$ interactions of users concentrate on the most popular items (most of them are also well-known and/or classic movies). However, the context of the user's last interactions (the last $15$ interactions in our case) becomes more personalized after multiple rounds of updates by the internal engine. Thus, the removal of the last $15$ interactions makes it more difficult to predict the test instance correctly, especially for sequential recommendation algorithms. From experiments with ItemKNN, SVD, and Multi-VAE, we can easily see that users' ratings of high-popularity items in the $R_1$ stage are of limited help to the detailed classification of users' preferences.

Further, we also observe the popularity dilemma posed by the first $15$ interactions. When the first $15$ interactions are included, the recommendation algorithm tends to treat the highly popular movies in the first $15$ interactions as the correct test items, which in turn affects the recommendation performance. When the first $15$ interactions are removed, the information about the user's primary interests embedded in the highly popular movies is lost, and the final result declines. 

To address the popularity dilemma of MovieLens, we note that \citet{pellegrini2022don} have proposed a modification: considering the co-occurrence probability of test samples with the most popular items as the objective function under popularity-sampled metrics. This inadvertently exploits group interest invariance in MovieLens: users' test items are essentially identical in group-based interest to the most popular items they rate in their first $15$ interactions. Thus, the probability of co-occurrence with the most popular items allows for the determination of the preference backbone to which test items are attributed.

\subsection{Impact of Interaction Sequence}
\label{ssec:intSequence}

\begin{figure}
    \centering
     \begin{subfigure}[t]{0.65\columnwidth}
        \includegraphics[width=\columnwidth]{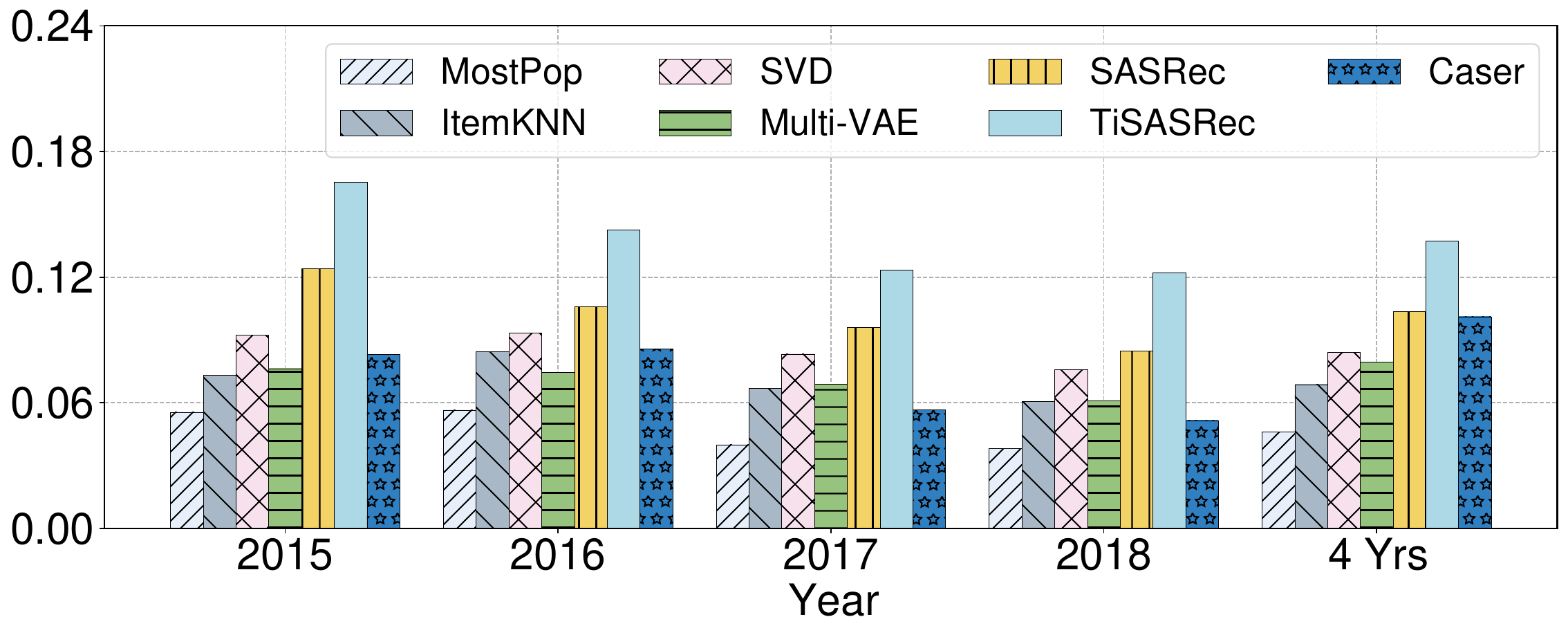}
        \caption{HR@10}
        \label{sfig:hr10}
    \end{subfigure}
  \begin{subfigure}[t]{0.65\columnwidth}
        \includegraphics[width=\columnwidth]{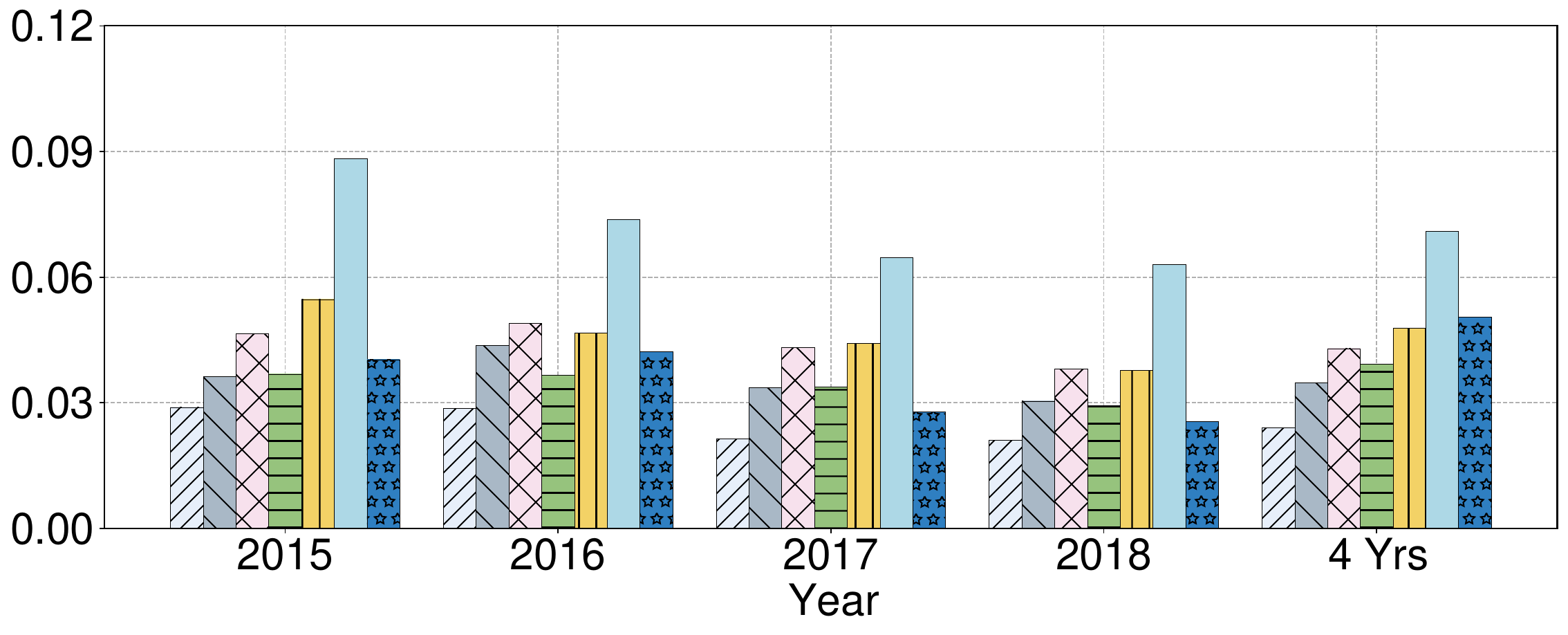}
        \caption{NDCG@10}
        \label{sfig:ndcg10}
    \end{subfigure}
        \caption{HR@10 and NDCG@10 results of seven baselines without data removal from training set, on four yearly datasets, and also the entire \dataset indicated by "4 Yrs".}
    \label{fig:comp_all_baselines}
\end{figure}

The plots in Figure~\ref{fig:allvsremove} do not provide a direct comparison of model performance. Figure~\ref{fig:comp_all_baselines} provides a direct comparison of the seven baselines without training data removal, measured by HR@10 and NDCG@10 respectively. Observe that TiSASRec is the best performer followed by SASRec, both are sequence-aware models. In general, there is a degradation trend for all models from the 2015 subset to 2018 subset.

Recall that movies rated by users in the MovieLens dataset are from a list of candidate movies recommended by an internal recommendation engine, in an iterative manner. Therefore, it is easy to see that user interactions under the MovieLens  interaction mechanism are highly sequential, following a latent sequence determined by the internal recommendation engine. On the other hand, there is no significant changes in user preference based on the study on the ratio of rated movies in the top 3 genres,  which denotes that user interaction sequences are based on a high degree of conformity to the MovieLens interaction generation mechanism.

As previously stated in Section~\ref{sec:datasetAnalysis}, MovieLens records the user's choices and evaluations based on his or her memory among the candidate items, rather than the user's actual movie-watching behavior. Driven by the interaction generation mechanism, the exposure of a movie at different stages of user interaction is not uniformly distributed, which in turn may introduce an implicit bias in the user sequence.

To further analyze the impact of potential biases introduced into the user interaction sequence by the interaction generation mechanism, our final experiment changes the order in the original sequence by data shuffling.  Again, we follow the leave-last-one-out scheme, and changes are only made on the training set for a fair comparison. Specifically, we keep the validation set and test set unchanged, get new pseudo-sequences by disrupting the order of user interaction sequences in the training set, and observe the performance changes of the sequence recommendation algorithm. We repeat the experiment three times with different seeds.

\begin{table}[t]
  \center
   \caption{The impact of data shuffling on the performance of sequential models on \dataset. We show the average algorithm performance over four years.}
  \label{tab:R3}
  \renewcommand\arraystretch{1.2}
  \begin{tabular}{l|c|c|c|c}
      \toprule
      Model &  Metric &  Original Sequence & Shuffled Sequence & \% decrease after shuffling
      \\
      \midrule
      \multirow{2}*{SASRec} & HR@10 & 0.1034  & 0.0604 & 41.61\% \\
		~ & NDCG@10 & 0.0464 & 0.0299 & 35.51\% \\
      \hline
     \multirow{2}*{TiSASRec} & HR@10 & 0.1378 & 0.0722 & 47.61\% \\
      ~ & NDCG@10 & 0.0687 & 0.0342 & 50.23\%\\
      \hline
      \multirow{2}*{Caser} & HR@10 & 0.0713 & 0.0342 & 52.00\% \\
      ~ & NDCG@10 & 0.0535 & 0.0271 & 49.44\%\\
      \hline
  \end{tabular}
 
\end{table}

\begin{observation}
After shuffling the data, the performance of sequential recommendation algorithms, which initially performed well, dropped significantly. The drop in performance of SASRec is slightly smaller, compared to TiSASRec and Caser.
\end{observation}

Our results are partially consistent with that reported in~\cite{woolridge2021sequence}, which suggests that the sequences in MovieLens are pseudo-sequences because a large part of SASRec's performance comes from modeling the internal recommendation engine of the MovieLens system rather than true sequence information~\cite{harper2015movielens}. The difference is that our results are based on full-rank evaluation while the results in~\cite{woolridge2021sequence} are based on sampled evaluation. We also include two more sequence-aware models TiSASRec and Caser. 

In terms of interaction contexts, data shuffling mixes candidate movies at different interaction stages. Actually, the overlap between the candidate movie pools in two adjacent interaction phases is less than 50\% (see Table~\ref{tab:ioufield}), which implies a significant difference between the different stages. Therefore, data shuffling increases the difficulty of interaction context simulation and accurate recommendations. Also, data shuffling destroys most of the potential information provided by the internal recommendation engine.

Based on all experiments so far, user interactions with MovieLens exhibit a consistent and significant pattern: the hierarchical expansion of the candidate item pool and the stability of group preference. Therefore, the efficacy of sequence-aware models on the MovieLens dataset primarily hinges on their capacity to comprehend this specific pattern. The method of sequence modeling, the various sequence stages, and the temporal distribution information collectively impact the final performance of a sequence-aware model to different degrees. The performance of the TiSASRec model is significantly enhanced by the inclusion of local temporal distribution modeling compared to SASRec, due to the more detailed modeling of the user interaction context. In addition, the performance of Caser, which is based on localized sequence information, is not stable enough on the smaller scale per-year dataset (see Figure~\ref{fig:comp_all_baselines}).

\section{Discussion}
\label{sec:discussion}

To summarize, what we have learned from the MovieLens user-item interaction generation mechanism and the \dataset dataset are the followings. (i) The movie ratings on MovieLens are collected from users through interactions at different stages. The movies for rating are recommended by an internal recommendation algorithm. Along the interactions at different stages, the receptive fields of users expand. (ii) Nearly half of the users complete all their ratings in a single day and more than 85\% of users complete all ratings within 5 days. (iii) Users likely maintain their preferences reflected by nearly 90\% of movies in the first 15 ratings fall into the top 3 genres of most interest, and above 70\% of last 15 ratings fall into the same genres. (iv) Removal of training instances (\eg the first 15, the last 15, and randomly sampled 15) leads to decreases in recommendation accuracy, and larger drops are observed for removal of training instances closer (\eg the last 15 ratings) to the test instances. (v) The best-performing model TiSASRec benefits from the sequential pattern of the internal recommendation, and data shuffle leads to a large drop in performance. (vi) All baseline models show a degradation trend on the yearly subsets from 2015 to 2018 due to the increase in movie candidates.

Our main focus here is how to explain these findings, and the implications of using the MovieLens dataset for model evaluation. In other words: \textit{are similar findings expected on a typical recommendation scenario in practice?} To answer this question, we focus on the similarities and differences between the user-item interaction generation mechanisms on MovieLens and on typical recommendation platforms, from two perspectives.

\subsection{The Two Types of User-Item Interactions}

\begin{figure}[t]
    \centering
    \includegraphics[trim = 3.2cm 12cm 14.5cm 5cm, clip, width = 0.7\linewidth]{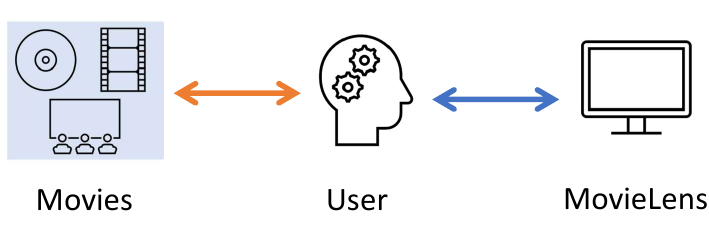}
    \caption{Illustration of two different kinds of interactions: The $\langle user - movie\rangle$ interactions may be accumulated over a long time period through different forms of interaction \eg movie watching in theater or on streaming platform;  The $\langle user - MovieLens\rangle$ interactions typically happen within a day and user rates movies that he/she ever watched.}
    \label{fig:twomode}
\end{figure}

In our understanding, we consider that the MovieLens platform is an effective platform for \textit{collecting} user preferences on movies (or similar items like books, music, video games, poetry, TV shows) at a large scale, \textit{for cold-start setting}. The website is well designed to provide a small (such as 50 movies in one selection page) but personalized list of titles for a user to choose from, as a "guide" to help him/her to \textit{recall the movies watched earlier} and then give ratings. There are two forms of user-item interactions, as illustrated in Figure~\ref{fig:twomode}. 
\begin{itemize}
    \item The $\langle user - movie \rangle$ interaction is between users and the individual movies which happened in the past, and is accumulative over time. A user may spend 10 years to complete the watching of Harry Potter series as these movies were released in a 10-year time frame. The decision-making to watch every Harry Potter movie could be different. In other words, a user may decide to watch a movie because of its actors and directors, simply because he/she receives a free ticket, or by following the recommendation on a platform, among many possible reasons. For each of such decisions, there could be consideration of the monetary and time cost of watching a movie. 
    \item The $\langle user - MovieLens\rangle$ interaction between the users and the MovieLens platform. Following the "guidance" of the MovieLens' internal recommendation, the user recalls the movies that he/she had watched before and records his/her ratings.
\end{itemize}

The two types of interactions are completely different. In particular, when interacting with the MovieLens platform, a user has watched a good number of movies. It is not difficult to \textit{summarize} the kinds of movies (\eg the top three genres) he/she likes. Hence, this rating collection process can be completed within a short time, and the process of recalling the movie titles is likely from the small pool of more famous movies (in stage $R_1$) to a much larger pool of movies of similar kinds (in stage $R_2$). There are also no preference changes over time as users usually complete all ratings within a short time based on their developed preferences on movies over the time before the interaction with MovieLens.  More importantly, the sequential pattern among the ratings from a user does not necessarily reflect the actual sequence of a user watching these movies~\cite{harper2015movielens,JannachZGG12}. Rather, the sequence is the result of the guidance through the rating and recommendation iterative process on MovieLens. Even if there were personal preference changes when the user watched these movies along time, such changes cannot be reflected in the rating sequence. 

As a user has already watched a movie, there is no decision-making process on whether a new movie shall be watched with either monetary, time or other forms of cost. However, for a typical online recommender platform, a user often needs to make a judgment before interacting with a "new" item being recommended, among the choices available at that decision time.  In other words, in a typical recommendation platform, we expect the model to recommend "new" items of the user's interest that the user has not interacted with before. The MovieLens dataset does not reflect such a setting. Again, we are not the first to question the usefulness of the MovieLens datasets from this perspective. In 2012, \citet{JannachZGG12} discussed that "the context
of movie consumption (Am I watching a movie alone or with friends? Am I
looking for entertainment or for intellectual challenge?) is largely not taken into account in the majority of papers" and the prediction on the dataset could be "accurate but not valuable." Nevertheless, to the best of our knowledge, our paper is the first attempt to formalize the two kinds of user-item interactions \ie  
the $\langle user - movie \rangle$ interactions and the $\langle user - MovieLens\rangle$ interactions, for a clearer comprehension of the dataset. In short, all research papers using the MovieLens dataset model the  $\langle user - MovieLens\rangle$ rather than the $\langle user - movie\rangle$ interactions, making their results less generalizable  to many practical recommendation scenarios in real-world settings. 

\citet{yusun2022reality} argue that  "a model is trained to answer the same information need in a similar context (\eg the information available), for which the training dataset is created". That is, a model learns user behaviour from the training data, with the assumption that similar user behaviour will happen when the model is applied in a different or real-world setting. In our discussion, we consider that the contexts for the  $\langle user - MovieLens\rangle$ interactions and the  $\langle user - movie \rangle$ interactions to be very different. Hence, a model that achieves excellent performance on MovieLens may not show superior performance in many practical recommendation scenarios, because there are many more factors that would affect the user's behavior in each decision making, but the decision-making  for the  $\langle user - MovieLens\rangle$ interaction is at nearly no cost. 

Evaluating RecSys models is challenging, particularly under an offline setting, because many factors that may affect user online behavior are not well captured in an offline dataset. \citet{Gunawardana2022} state that "the goal of the offline experiments is to filter out inappropriate approaches, leaving a relatively small set of candidate algorithms to be tested" online, and "it is necessary to simulate the online process where the system makes predictions or recommendations". As illustrated in Figure~\ref{fig:twomode}, the kind of $\langle user-MovieLens\rangle$ interactions collected in the MovieLens dataset cannot be used to simulate an online process for users to make judgements on what movies to watch next \ie $\langle user-movie\rangle$ interactions. Although the recommended usage of the MovieLens dataset is for algorithm comparison in research papers~\cite{harper2015movielens}, we hope that through our analysis, researchers have a better understanding of the results obtained on the MovieLens dataset.

\subsection{The Information Mismatch between User and Item Collection}

In MovieLens, both users and the system have a clear understanding of the collection of items, 
\ie the movies. A typical user would have a basic comprehension of how to characterize a movie with its actors, directors, genre and so on.  The same applies to music, books, hotels and flights. However, in many other recommendation scenarios, users may not have a clear understanding of the organization and distribution of all items, \eg all products available for purchase on e-commerce sites, and the different kinds of advertisements to be recommended to users. Under such settings, users may not know the existence of certain types of items and do not have a full picture of all possible types of items that are available for recommendation. The degree of information mismatch between the user and the item collection of the MovieLens dataset and that of other recommendation scenarios could be very different. Hence, a model that is good at learning certain patterns from the MovieLens dataset may not be good at learning patterns from datasets obtained from other platforms. 

In short, there are two perspectives to understanding the MovieLens data. First, the MovieLens dataset does capture user preferences on movies \eg the types of movies a user prefers to watch, from all movies that a user has watched so far. Second, the MovieLens dataset is more of a collection of the \textit{results} of user-item interactions, where users have good knowledge about the item properties. Hence, the model performance made on the MovieLens dataset could be a good reflection of to what extent the model matches the underlying recommendation algorithm well (\eg item-item similarity used to power MovieLens); the performance may not be a good reflection of what the model would achieve in real-world settings where a user need to make a judgment when facing newly recommended items of little knowledge, and the user has never interacted with these items before, and there could be cost incurred to interact with these items.

\subsection{Information Collection for Cold Starts}

Cold start is one of the major challenges in many recommender systems, where the system does not know much about the new users (\ie user cold start) and/or the new items (\ie item cold start). Preference elicitation in the cold-start phase determines the quality of subsequent recommendations and thus has been studied extensively. Researchers strive to propose new preference elicitation methods to better address the cold-start problem. For instance,  \citet{sepliarskaia2018preference} define preference elicitation as an optimization problem to generate \textit{static preference questionnaires} at a lower cost. \citet{graus2015improving} argue that the choice-based interface requires less user effort than rating-based interfaces, and leads to a more satisfying recommendation experience.

The MovieLens platform in this perspective demonstrates an efficient and effective way of collecting user preferences in a personalized iterative process. In particular, it starts with a group-based preference elicitation, and then guides the users to "recall" the movies that they have interacted with through an recommendation process. As the result, for a new user, the platform is able to collect a good number of movie ratings within a very short time (\eg within a day for nearly half of all users), and with very little effort from users. These ratings collected well reflect the user preferences, which is demonstrated by the conformity to the top 3 genres, serving an effective way to understand users. In this sense, the MovieLens dataset can be considered as the result of static preference questionnaire collected through a well designed functional web interface. We use the term "static preference" to describe the user preference collected in MovieLens, for the reason that the user preference here is inferred from all ratings submitted within a short period. In other words, the user preference is a summary of all the movies he/she has watched over a much longer time period in no particular order. As there is no time point when a user watches which movie, no preference change over time can be inferred.

\subsection{Is It a Good Idea to Evaluate RecSys Models on MovieLens?}
The succinct answer is no.   Our analysis indicates that the user-item interactions recorded in the MovieLens dataset represent engagements between users and the MovieLens platform, where users are guided to recall movies they have previously watched. We perceive this setup as significantly divergent from typical recommendation scenarios encountered in practice. Notably, the MovieLens dataset emerges as a distinct outlier in the distribution of certain statistical features, as previously observed in studies~\cite{chin2022datasets,tang2018personalized}.  We believe that performances obtained on the MovieLens dataset do not adequately reflect a model's expected performance in real-world scenarios. Therefore, results obtained solely from the MovieLens dataset cannot be relied upon as indicative of a model's online performance post-deployment.
    
On the other hand, MovieLens stands out as one of the most popular datasets in the field of recommender systems~\cite{chin2022datasets, sun2022daisyrec}. Results derived from MovieLens serve as valuable references for researchers to validate their implementations. For instance, researchers can compare the outcomes of their own implementations with the results reported by the authors on the MovieLens dataset, as in Appendix~\ref{sec:appdx}, Table~\ref{tab:evaML1M}.

In short, while providing results on MovieLens for reference purposes is beneficial, it should not serve as a strong justification for the effectiveness of a proposed model. Models should be evaluated on a variety of datasets, not relying solely on the MovieLens dataset.

\section{Related Work}
\label{sec:related}

Our research focuses on the analysis of the MovieLens dataset with an aim to identify its implications on recommender evaluations. Accordingly, we review the related studies on RecSys dataset analysis, the study of biases in RecSys, and the common issues in RecSys evaluation. 

\subsection{Recommender System Dataset Analysis}

Existing dataset related studies in recommender systems have placed their focuses on dataset analysis.~\citet{deldjoo2020dataset} emphasize that the user-item rating matrix structure and the rating distribution largely affect the robustness of the collaborative filtering model in the case of shilling attacks. \citet{luca2016reviews} examine the causal impact of Yelp consumer ratings on restaurant demand with a regression discontinuity approach. The authors then test whether consumers use Yelp reviews in a way that is consistent with standard Bayesian learning models. The results suggest that consumers show selectivity in using reviews, being more responsive to visible quality changes and ratings with more information.~\citet{leino2007case} explore user behaviors in the Amazon online store using on-location interviewing and observation. They are especially interested in what kind of strategies users had developed for utilizing algorithm-based recommendations and customer reviews to discover items of interest.~\citet{steck2021deep} discuss the challenges and lessons learned in using deep learning for recommender systems at Netflix. It highlights that deep learning models show significant improvements only when combined with additional heterogeneous features like timestamps and by addressing issues of offline-online metric alignment. In~\cite{woolridge2021sequence}, the authors demonstrate that the interaction history in the MovieLens dataset is pseudo-sequential because the actual order of these interactions is unknowable. Similar to our study, these papers mostly focus on one specific dataset. Considering the arbitrary use of various datasets, the dataset dilemma in recommender systems has been raised in~\cite{chin2022datasets}. The authors cluster datasets from different domains based on structural and distributional characteristics, and examine the performance rankings of algorithms across different clusters.

\subsection{Biases in Recommender Systems}

The recommendation system can be abstracted as a feedback loop among three key components: User, Data, and Model~\cite{chen2023bias}. Biases occur in different stages including data collection (User to Data), model learning (Data to Model), and final recommendation (Model to User). Our study pays more attention to biases in collected data and recommendation results. Specifically, \citet{chen2017common} formally present four common pitfalls in training and evaluating recommendation algorithms. Among them, two pitfalls are highly relevant to this study: (1) 
popularity bias: trained models could be biased toward highly reachable products because these items are more likely to be treated as positive training instances. 
(2) exposure bias: interactions collected from an online platform are influenced by its deployed recommender system. Hence, the distribution of interactions is fundamentally different from the interactions with no exposure to a recommender system.

The long-tail distribution of user interactions is frequently observed in recommendation system data, where a small portion of highly popular items tends to dominate user interactions. When trained on such data, models often have the propensity to recommend popular items over less unpopular ones. MovieLens datasets, as a typical long-tail dataset, is often used as an experimental dataset for popularity bias and debiasing. \citet{abdollahpouri2020connection} have empirically verified the impacts of popularity bias on different stakeholders such as users and suppliers. Further, they propose metrics to measure the average deviation of the recommendations in terms of item popularity. \citet{steck2021deep} investigate the effects of four factors, namely inherent audience size imbalance, model bias, position bias, and closed feedback loop, on popularity bias in recommender systems by simulating dynamic recommendation experiments. They conclude that audience size imbalance and model bias are the main factors contributing to popularity bias. Apart from this, a dynamic debiasing strategy and a novel False Positive Correction method are proposed to remove popularity bias in dynamic scenarios. Exposure bias arises when users have only seen a subset of specific items and unobserved interactions are not necessarily dislikes.~\citet{liu2020general} point out that exposure is affected by the policy of intrinsic recommendation engines, which determine which items to show to users. \citet{schnabel2016recommendations} emphasize that users' active search behavior is also a factor of exposure and introduce a debiasing method employing inverse propensity weighting to correct bias in observed data. 
 
Our analysis in this paper offers another perspective to understand bias in the MovieLens dataset. The ratings on MovieLens are naturally biased due to two reasons. First, all users start with the 15 ratings from a small pool of representative movies, and the number of ratings per user varies significantly. Second, as the ratings are based on user memory of the movies that they have watched before, more ratings are expected to  more successful movies. 

\subsection{Evaluation Issues in Recommender Systems}

In the burgeoning field of recommender systems, evaluation has been an important research topic. In~\cite{ferrari2019we} and~\cite{newerNotBetter}, the authors highlight the importance of establishing benchmark evaluation to avoid unfair comparisons and to ensure the reproducibility of recommendation algorithms. With that in consideration, multiple toolkits have been released to benchmark the recommendation tasks, and facilitate the development and evaluation of recommendation models~\cite{sun2022daisyrec, zhao2021recbole, microsoftRecommender, Elliot}.

Apart from the development of toolkits, recent years have witnessed an increasing number of research papers discussing offline evaluation options for recommender systems.~\citet{sun2020we} categorize the offline evaluation options into eight steps, including dataset selection, dataset filtering techniques, baseline model selection, objective function selection, negative sampling method selection, dataset splitting options, evaluation metric options and hyperparameter tuning techniques. Among these options, \textit{dataset splitting options} have garnered significant research interests. In~\cite{canamares2020offline, meng2020exploring}, the authors demonstrate the varying outcomes due to different data splitting strategies. They argue that varying outcomes could lead to misinterpretation of results.~\citet{ji2023critical} further explain the varying outcomes from a global timeline perspective. The key finding is, the data splitting strategies that ignore the global timeline, \eg leave-one-out split, suffer from the data leakage issue, thus leading to unpredictable recommendation performance. Other studies focus on the options of \textit{evaluation metric}: accuracy metric and beyond-accuracy metric~\cite{evaluateRecsysSurvey, howGoodIsRecSys, evalCFModel}. Furthermore, the authors in~\cite{zhao2022revisiting,krichene2022sampled} delve into the research of sampled metric, and show the potential bias induced when using incomplete candidate list in evaluation.

Although \textit{dataset selection} has been listed as an important step of offline evaluation in~\cite{sun2020we}, it does not receive much research attention compared to \textit{data splitting strategies} and \textit{evaluation metrics}. In many studies~\cite{sun2020we, zhao2022revisiting}, it is suggested that researchers should select research datasets considering their popularity (whether they are frequently used by recent academic papers) as well as domain diversity. There is a lack of study that delves into the collection and construction process of a dataset to facilitate the dataset selection in offline evaluation. In this work, we zoom in on a particular recommender system dataset: MovieLens. Specifically, we understand the MovieLens dataset from its collection process.

\section{Conclusion}
\label{sec:conclude}

In this study, we conduct a thorough analysis of the extremely popular MovieLens dataset in the field of recommender systems from the perspective of interaction generation mechanisms. We demonstrate the interaction generation mechanism of the latest version of the MovieLens dataset, including the detailed user interaction process and the built-in recommendation algorithm at different stages. In addition, we designed targeted experiments based on the interactive generation mechanism to observe whether the special data characteristics caused by the interaction generation mechanism of the MovieLens dataset actually lead to different observations or conclusions drawn from empirical studies. Our results demonstrate that interaction generation mechanisms can have a significant impact on data characteristics, which in turn can cause profound and unpredictable perturbations to experimental results. 

The recommendation generation mechanism of the MovieLens platform notably impacts the data dynamics of user-platform interactions, thereby facilitating certain models to achieve superior performance as anticipated. There is no doubt that the MovieLens platform demonstrates the best way to collect user preferences with minimal effort from users and thus address cold-start problems. On the other hand, we argue that the recommendation models that achieve excellent performance on the MovieLens dataset may not show the same in reality by contrasting the interaction generation mechanism in MovieLens and other practical recommendation scenarios. Results on the MovieLens dataset hence should not
serve as a strong justification for the effectiveness of a proposed model. 

Our analysis is limited to the MovieLens dataset, one of the most widely used datasets in RecSys research. While not all findings reported in this paper are new, we hope to provide a comprehensive understanding of the dataset from the perspective of user-item generation mechanism. The findings will be helpful to researchers, particularly the young researchers who are new to RecSys, to better interpret the results of their models on benchmark datasets. While the methodology used to analyze the MovieLens dataset may not be applicable to all other RecSys datasets, we strongly encourage researchers to review the mechanisms by which the user-item interactions were generated when using datasets as benchmarks for recommenders beyond their initial scope.


\bibliographystyle{ACM-Reference-Format}

\bibliography{DataAnalysis}

\begin{thebibliography}{49}


\ifx \showCODEN    \undefined \def \showCODEN     #1{\unskip}     \fi
\ifx \showDOI      \undefined \def \showDOI       #1{#1}\fi
\ifx \showISBNx    \undefined \def \showISBNx     #1{\unskip}     \fi
\ifx \showISBNxiii \undefined \def \showISBNxiii  #1{\unskip}     \fi
\ifx \showISSN     \undefined \def \showISSN      #1{\unskip}     \fi
\ifx \showLCCN     \undefined \def \showLCCN      #1{\unskip}     \fi
\ifx \shownote     \undefined \def \shownote      #1{#1}          \fi
\ifx \showarticletitle \undefined \def \showarticletitle #1{#1}   \fi
\ifx \showURL      \undefined \def \showURL       {\relax}        \fi
\providecommand\bibfield[2]{#2}
\providecommand\bibinfo[2]{#2}
\providecommand\natexlab[1]{#1}
\providecommand\showeprint[2][]{arXiv:#2}

\bibitem[Abdollahpouri et~al\mbox{.}(2020)]%
        {abdollahpouri2020connection}
\bibfield{author}{\bibinfo{person}{Himan Abdollahpouri}, \bibinfo{person}{Masoud Mansoury}, \bibinfo{person}{Robin Burke}, {and} \bibinfo{person}{Bamshad Mobasher}.} \bibinfo{year}{2020}\natexlab{}.
\newblock \showarticletitle{The connection between popularity bias, calibration, and fairness in recommendation}. In \bibinfo{booktitle}{\emph{ACM Conference on Recommender Systems}}. \bibinfo{pages}{726--731}.
\newblock


\bibitem[Anelli et~al\mbox{.}(2021)]%
        {Elliot}
\bibfield{author}{\bibinfo{person}{Vito~Walter Anelli}, \bibinfo{person}{Alejandro Bellog{\'{\i}}n}, \bibinfo{person}{Antonio Ferrara}, \bibinfo{person}{Daniele Malitesta}, \bibinfo{person}{Felice~Antonio Merra}, \bibinfo{person}{Claudio Pomo}, \bibinfo{person}{Francesco~Maria Donini}, {and} \bibinfo{person}{Tommaso~Di Noia}.} \bibinfo{year}{2021}\natexlab{}.
\newblock \showarticletitle{Elliot: {A} Comprehensive and Rigorous Framework for Reproducible Recommender Systems Evaluation}. In \bibinfo{booktitle}{\emph{{ACM} {SIGIR} Conference on Research and Development in Information Retrieval}}. \bibinfo{publisher}{{ACM}}, \bibinfo{pages}{2405--2414}.
\newblock
\urldef\tempurl%
\url{https://doi.org/10.1145/3404835.3463245}
\showDOI{\tempurl}


\bibitem[Ca{\~n}amares et~al\mbox{.}(2020)]%
        {canamares2020offline}
\bibfield{author}{\bibinfo{person}{Roc{\'\i}o Ca{\~n}amares}, \bibinfo{person}{Pablo Castells}, {and} \bibinfo{person}{Alistair Moffat}.} \bibinfo{year}{2020}\natexlab{}.
\newblock \showarticletitle{Offline evaluation options for recommender systems}.
\newblock \bibinfo{journal}{\emph{Information Retrieval Journal}} \bibinfo{volume}{23}, \bibinfo{number}{4} (\bibinfo{year}{2020}), \bibinfo{pages}{387--410}.
\newblock


\bibitem[Chang et~al\mbox{.}(2015)]%
        {chang2015using}
\bibfield{author}{\bibinfo{person}{Shuo Chang}, \bibinfo{person}{F~Maxwell Harper}, {and} \bibinfo{person}{Loren Terveen}.} \bibinfo{year}{2015}\natexlab{}.
\newblock \showarticletitle{Using groups of items for preference elicitation in recommender systems}. In \bibinfo{booktitle}{\emph{ACM Conference on Computer Supported Cooperative Work \& Social Computing}}. \bibinfo{pages}{1258--1269}.
\newblock


\bibitem[Chen et~al\mbox{.}(2017)]%
        {chen2017common}
\bibfield{author}{\bibinfo{person}{Hung-Hsuan Chen}, \bibinfo{person}{Chu-An Chung}, \bibinfo{person}{Hsin-Chien Huang}, {and} \bibinfo{person}{Wen Tsui}.} \bibinfo{year}{2017}\natexlab{}.
\newblock \showarticletitle{Common pitfalls in training and evaluating recommender systems}.
\newblock \bibinfo{journal}{\emph{ACM SIGKDD Explorations Newsletter}} \bibinfo{volume}{19}, \bibinfo{number}{1} (\bibinfo{year}{2017}), \bibinfo{pages}{37--45}.
\newblock


\bibitem[Chen et~al\mbox{.}(2023)]%
        {chen2023bias}
\bibfield{author}{\bibinfo{person}{Jiawei Chen}, \bibinfo{person}{Hande Dong}, \bibinfo{person}{Xiang Wang}, \bibinfo{person}{Fuli Feng}, \bibinfo{person}{Meng Wang}, {and} \bibinfo{person}{Xiangnan He}.} \bibinfo{year}{2023}\natexlab{}.
\newblock \showarticletitle{Bias and debias in recommender system: A survey and future directions}.
\newblock \bibinfo{journal}{\emph{ACM Transactions on Information Systems}} \bibinfo{volume}{41}, \bibinfo{number}{3} (\bibinfo{year}{2023}), \bibinfo{pages}{1--39}.
\newblock


\bibitem[Chin et~al\mbox{.}(2022)]%
        {chin2022datasets}
\bibfield{author}{\bibinfo{person}{Jin~Yao Chin}, \bibinfo{person}{Yile Chen}, {and} \bibinfo{person}{Gao Cong}.} \bibinfo{year}{2022}\natexlab{}.
\newblock \showarticletitle{The Datasets Dilemma: How Much Do We Really Know About Recommendation Datasets?}. In \bibinfo{booktitle}{\emph{ACM International Conference on Web Search and Data Mining}}. \bibinfo{pages}{141--149}.
\newblock


\bibitem[Cremonesi et~al\mbox{.}(2010)]%
        {DBLP:conf/recsys/CremonesiKT10}
\bibfield{author}{\bibinfo{person}{Paolo Cremonesi}, \bibinfo{person}{Yehuda Koren}, {and} \bibinfo{person}{Roberto Turrin}.} \bibinfo{year}{2010}\natexlab{}.
\newblock \showarticletitle{Performance of recommender algorithms on top-n recommendation tasks}. In \bibinfo{booktitle}{\emph{{ACM} Conference on Recommender Systems}}, \bibfield{editor}{\bibinfo{person}{Xavier Amatriain}, \bibinfo{person}{Marc Torrens}, \bibinfo{person}{Paul Resnick}, {and} \bibinfo{person}{Markus Zanker}} (Eds.). \bibinfo{publisher}{{ACM}}, \bibinfo{pages}{39--46}.
\newblock
\urldef\tempurl%
\url{https://doi.org/10.1145/1864708.1864721}
\showDOI{\tempurl}


\bibitem[Deldjoo et~al\mbox{.}(2020)]%
        {deldjoo2020dataset}
\bibfield{author}{\bibinfo{person}{Yashar Deldjoo}, \bibinfo{person}{Tommaso Di~Noia}, \bibinfo{person}{Eugenio Di~Sciascio}, {and} \bibinfo{person}{Felice~Antonio Merra}.} \bibinfo{year}{2020}\natexlab{}.
\newblock \showarticletitle{How dataset characteristics affect the robustness of collaborative recommendation models}. In \bibinfo{booktitle}{\emph{ACM SIGIR conference on research and development in information retrieval}}. \bibinfo{pages}{951--960}.
\newblock


\bibitem[Deshpande and Karypis(2004)]%
        {DBLP:journals/tois/DeshpandeK04}
\bibfield{author}{\bibinfo{person}{Mukund Deshpande} {and} \bibinfo{person}{George Karypis}.} \bibinfo{year}{2004}\natexlab{}.
\newblock \showarticletitle{Item-based top-\emph{N} recommendation algorithms}.
\newblock \bibinfo{journal}{\emph{{ACM} Trans. Inf. Syst.}} \bibinfo{volume}{22}, \bibinfo{number}{1} (\bibinfo{year}{2004}), \bibinfo{pages}{143--177}.
\newblock
\urldef\tempurl%
\url{https://doi.org/10.1145/963770.963776}
\showDOI{\tempurl}


\bibitem[Dong et~al\mbox{.}(2023)]%
        {newerNotBetter}
\bibfield{author}{\bibinfo{person}{Yushun Dong}, \bibinfo{person}{Jundong Li}, {and} \bibinfo{person}{Tobias Schnabel}.} \bibinfo{year}{2023}\natexlab{}.
\newblock \showarticletitle{When Newer is Not Better: Does Deep Learning Really Benefit Recommendation From Implicit Feedback?}
\newblock \bibinfo{journal}{\emph{CoRR}}  \bibinfo{volume}{abs/2305.01801} (\bibinfo{year}{2023}).
\newblock
\urldef\tempurl%
\url{https://doi.org/10.48550/arXiv.2305.01801}
\showDOI{\tempurl}
\showeprint[arXiv]{2305.01801}


\bibitem[Ekstrand et~al\mbox{.}(2015)]%
        {ekstrand2015letting}
\bibfield{author}{\bibinfo{person}{Michael~D Ekstrand}, \bibinfo{person}{Daniel Kluver}, \bibinfo{person}{F~Maxwell Harper}, {and} \bibinfo{person}{Joseph~A Konstan}.} \bibinfo{year}{2015}\natexlab{}.
\newblock \showarticletitle{Letting users choose recommender algorithms: An experimental study}. In \bibinfo{booktitle}{\emph{Proceedings of the 9th ACM Conference on Recommender Systems}}. \bibinfo{pages}{11--18}.
\newblock


\bibitem[Ferrari~Dacrema et~al\mbox{.}(2019)]%
        {ferrari2019we}
\bibfield{author}{\bibinfo{person}{Maurizio Ferrari~Dacrema}, \bibinfo{person}{Paolo Cremonesi}, {and} \bibinfo{person}{Dietmar Jannach}.} \bibinfo{year}{2019}\natexlab{}.
\newblock \showarticletitle{Are we really making much progress? A worrying analysis of recent neural recommendation approaches}. In \bibinfo{booktitle}{\emph{ACM conference on recommender systems}}. \bibinfo{publisher}{Association for Computing Machinery}, \bibinfo{address}{New York, NY, USA}, \bibinfo{pages}{101--109}.
\newblock


\bibitem[Graham et~al\mbox{.}(2019)]%
        {microsoftRecommender}
\bibfield{author}{\bibinfo{person}{Scott Graham}, \bibinfo{person}{Jun{-}Ki Min}, {and} \bibinfo{person}{Tao Wu}.} \bibinfo{year}{2019}\natexlab{}.
\newblock \showarticletitle{Microsoft recommenders: tools to accelerate developing recommender systems}. In \bibinfo{booktitle}{\emph{{ACM} Conference on Recommender Systems}}. \bibinfo{publisher}{{ACM}}, \bibinfo{pages}{542--543}.
\newblock
\urldef\tempurl%
\url{https://doi.org/10.1145/3298689.3346967}
\showDOI{\tempurl}


\bibitem[Graus and Willemsen(2015)]%
        {graus2015improving}
\bibfield{author}{\bibinfo{person}{Mark~P Graus} {and} \bibinfo{person}{Martijn~C Willemsen}.} \bibinfo{year}{2015}\natexlab{}.
\newblock \showarticletitle{Improving the user experience during cold start through choice-based preference elicitation}. In \bibinfo{booktitle}{\emph{Proceedings of the 9th ACM Conference on Recommender Systems}}. \bibinfo{pages}{273--276}.
\newblock


\bibitem[Gunawardana et~al\mbox{.}(2022)]%
        {Gunawardana2022}
\bibfield{author}{\bibinfo{person}{Asela Gunawardana}, \bibinfo{person}{Guy Shani}, {and} \bibinfo{person}{Sivan Yogev}.} \bibinfo{year}{2022}\natexlab{}.
\newblock \showarticletitle{Evaluating Recommender Systems}.
\newblock In \bibinfo{booktitle}{\emph{Recommender Systems Handbook (3rd Ed)}}, \bibfield{editor}{\bibinfo{person}{Francesco Ricci}, \bibinfo{person}{Lior Rokach}, {and} \bibinfo{person}{Bracha Shapira}} (Eds.). \bibinfo{publisher}{Springer US}, \bibinfo{address}{New York, NY}, \bibinfo{pages}{547--601}.
\newblock
\urldef\tempurl%
\url{https://doi.org/10.1007/978-1-0716-2197-4_15}
\showDOI{\tempurl}


\bibitem[Harper and Konstan(2015)]%
        {harper2015movielens}
\bibfield{author}{\bibinfo{person}{F~Maxwell Harper} {and} \bibinfo{person}{Joseph~A Konstan}.} \bibinfo{year}{2015}\natexlab{}.
\newblock \showarticletitle{The movielens datasets: History and context}.
\newblock \bibinfo{journal}{\emph{ACM transactions on interactive intelligent systems}} \bibinfo{volume}{5}, \bibinfo{number}{4} (\bibinfo{year}{2015}), \bibinfo{pages}{1--19}.
\newblock


\bibitem[He et~al\mbox{.}(2017)]%
        {he2017neural}
\bibfield{author}{\bibinfo{person}{Xiangnan He}, \bibinfo{person}{Lizi Liao}, \bibinfo{person}{Hanwang Zhang}, \bibinfo{person}{Liqiang Nie}, \bibinfo{person}{Xia Hu}, {and} \bibinfo{person}{Tat-Seng Chua}.} \bibinfo{year}{2017}\natexlab{}.
\newblock \showarticletitle{Neural collaborative filtering}. In \bibinfo{booktitle}{\emph{International conference on world wide web}}. \bibinfo{publisher}{International World Wide Web Conferences Steering Committee}, \bibinfo{address}{Republic and Canton of Geneva, CHE}, \bibinfo{pages}{173--182}.
\newblock


\bibitem[Herlocker et~al\mbox{.}(2004)]%
        {evalCFModel}
\bibfield{author}{\bibinfo{person}{Jonathan~L. Herlocker}, \bibinfo{person}{Joseph~A. Konstan}, \bibinfo{person}{Loren~G. Terveen}, {and} \bibinfo{person}{John Riedl}.} \bibinfo{year}{2004}\natexlab{}.
\newblock \showarticletitle{Evaluating collaborative filtering recommender systems}.
\newblock \bibinfo{journal}{\emph{{ACM} Trans. Inf. Syst.}} \bibinfo{volume}{22}, \bibinfo{number}{1} (\bibinfo{year}{2004}), \bibinfo{pages}{5--53}.
\newblock
\urldef\tempurl%
\url{https://doi.org/10.1145/963770.963772}
\showDOI{\tempurl}


\bibitem[Hidasi and Czapp(2023)]%
        {thirdPartyImplementation}
\bibfield{author}{\bibinfo{person}{Bal{\'{a}}zs Hidasi} {and} \bibinfo{person}{{\'{A}}d{\'{a}}m~Tibor Czapp}.} \bibinfo{year}{2023}\natexlab{}.
\newblock \showarticletitle{The Effect of Third Party Implementations on Reproducibility}. In \bibinfo{booktitle}{\emph{Proceedings of the 17th {ACM} Conference on Recommender Systems, RecSys 2023, Singapore, Singapore, September 18-22, 2023}}, \bibfield{editor}{\bibinfo{person}{Jie Zhang}, \bibinfo{person}{Li~Chen}, \bibinfo{person}{Shlomo Berkovsky}, \bibinfo{person}{Min Zhang}, \bibinfo{person}{Tommaso~Di Noia}, \bibinfo{person}{Justin Basilico}, \bibinfo{person}{Luiz Pizzato}, {and} \bibinfo{person}{Yang Song}} (Eds.). \bibinfo{publisher}{{ACM}}, \bibinfo{pages}{272--282}.
\newblock
\urldef\tempurl%
\url{https://doi.org/10.1145/3604915.3609487}
\showDOI{\tempurl}


\bibitem[Hidasi et~al\mbox{.}(2016)]%
        {DBLP:journals/corr/HidasiKBT15}
\bibfield{author}{\bibinfo{person}{Bal{\'{a}}zs Hidasi}, \bibinfo{person}{Alexandros Karatzoglou}, \bibinfo{person}{Linas Baltrunas}, {and} \bibinfo{person}{Domonkos Tikk}.} \bibinfo{year}{2016}\natexlab{}.
\newblock \showarticletitle{Session-based Recommendations with Recurrent Neural Networks}. In \bibinfo{booktitle}{\emph{International Conference on Learning Representations}}, \bibfield{editor}{\bibinfo{person}{Yoshua Bengio} {and} \bibinfo{person}{Yann LeCun}} (Eds.).
\newblock


\bibitem[Jannach et~al\mbox{.}(2012)]%
        {JannachZGG12}
\bibfield{author}{\bibinfo{person}{Dietmar Jannach}, \bibinfo{person}{Markus Zanker}, \bibinfo{person}{Mouzhi Ge}, {and} \bibinfo{person}{Marian Gr{\"{o}}ning}.} \bibinfo{year}{2012}\natexlab{}.
\newblock \showarticletitle{Recommender Systems in Computer Science and Information Systems - {A} Landscape of Research}. In \bibinfo{booktitle}{\emph{Proc. E-Commerce and Web Technologies (EC-Web)}} \emph{(\bibinfo{series}{Lecture Notes in Business Information Processing}, Vol.~\bibinfo{volume}{123})}. \bibinfo{publisher}{Springer}, \bibinfo{pages}{76--87}.
\newblock
\urldef\tempurl%
\url{https://doi.org/10.1007/978-3-642-32273-0\_7}
\showDOI{\tempurl}


\bibitem[Ji et~al\mbox{.}(2023)]%
        {ji2023critical}
\bibfield{author}{\bibinfo{person}{Yitong Ji}, \bibinfo{person}{Aixin Sun}, \bibinfo{person}{Jie Zhang}, {and} \bibinfo{person}{Chenliang Li}.} \bibinfo{year}{2023}\natexlab{}.
\newblock \showarticletitle{A critical study on data leakage in recommender system offline evaluation}.
\newblock \bibinfo{journal}{\emph{ACM Transactions on Information Systems}} \bibinfo{volume}{41}, \bibinfo{number}{3} (\bibinfo{year}{2023}), \bibinfo{pages}{1--27}.
\newblock


\bibitem[Kang and McAuley(2018a)]%
        {DBLP:conf/icdm/KangM18}
\bibfield{author}{\bibinfo{person}{Wang{-}Cheng Kang} {and} \bibinfo{person}{Julian~J. McAuley}.} \bibinfo{year}{2018}\natexlab{a}.
\newblock \showarticletitle{Self-Attentive Sequential Recommendation}. In \bibinfo{booktitle}{\emph{{IEEE} International Conference on Data Mining, {ICDM}}}. \bibinfo{publisher}{{IEEE} Computer Society}, \bibinfo{pages}{197--206}.
\newblock
\urldef\tempurl%
\url{https://doi.org/10.1109/ICDM.2018.00035}
\showDOI{\tempurl}


\bibitem[Kang and McAuley(2018b)]%
        {kang2018self}
\bibfield{author}{\bibinfo{person}{Wang-Cheng Kang} {and} \bibinfo{person}{Julian McAuley}.} \bibinfo{year}{2018}\natexlab{b}.
\newblock \showarticletitle{Self-attentive sequential recommendation}. In \bibinfo{booktitle}{\emph{IEEE international conference on data mining (ICDM)}}. IEEE, \bibinfo{pages}{197--206}.
\newblock


\bibitem[Klimashevskaia et~al\mbox{.}(2023)]%
        {klimashevskaia2023survey}
\bibfield{author}{\bibinfo{person}{Anastasiia Klimashevskaia}, \bibinfo{person}{Dietmar Jannach}, \bibinfo{person}{Mehdi Elahi}, {and} \bibinfo{person}{Christoph Trattner}.} \bibinfo{year}{2023}\natexlab{}.
\newblock \bibinfo{title}{A Survey on Popularity Bias in Recommender Systems}.
\newblock
\newblock
\showeprint[arxiv]{2308.01118}~[cs.IR]


\bibitem[Krichene and Rendle(2022)]%
        {krichene2022sampled}
\bibfield{author}{\bibinfo{person}{Walid Krichene} {and} \bibinfo{person}{Steffen Rendle}.} \bibinfo{year}{2022}\natexlab{}.
\newblock \showarticletitle{On sampled metrics for item recommendation}.
\newblock \bibinfo{journal}{\emph{Commun. ACM}} \bibinfo{volume}{65}, \bibinfo{number}{7} (\bibinfo{year}{2022}), \bibinfo{pages}{75--83}.
\newblock


\bibitem[Leino and R{\"a}ih{\"a}(2007)]%
        {leino2007case}
\bibfield{author}{\bibinfo{person}{Juha Leino} {and} \bibinfo{person}{Kari-Jouko R{\"a}ih{\"a}}.} \bibinfo{year}{2007}\natexlab{}.
\newblock \showarticletitle{Case amazon: ratings and reviews as part of recommendations}. In \bibinfo{booktitle}{\emph{ACM conference on Recommender systems}}. \bibinfo{pages}{137--140}.
\newblock


\bibitem[Li et~al\mbox{.}(2020)]%
        {li2020time}
\bibfield{author}{\bibinfo{person}{Jiacheng Li}, \bibinfo{person}{Yujie Wang}, {and} \bibinfo{person}{Julian McAuley}.} \bibinfo{year}{2020}\natexlab{}.
\newblock \showarticletitle{Time interval aware self-attention for sequential recommendation}. In \bibinfo{booktitle}{\emph{Proceedings of the 13th international conference on web search and data mining}}. \bibinfo{pages}{322--330}.
\newblock


\bibitem[Liang et~al\mbox{.}(2018)]%
        {liang2018variational}
\bibfield{author}{\bibinfo{person}{Dawen Liang}, \bibinfo{person}{Rahul~G Krishnan}, \bibinfo{person}{Matthew~D Hoffman}, {and} \bibinfo{person}{Tony Jebara}.} \bibinfo{year}{2018}\natexlab{}.
\newblock \showarticletitle{Variational autoencoders for collaborative filtering}. In \bibinfo{booktitle}{\emph{The World Wide Web Conference}}. \bibinfo{pages}{689--698}.
\newblock


\bibitem[Liu et~al\mbox{.}(2020)]%
        {liu2020general}
\bibfield{author}{\bibinfo{person}{Dugang Liu}, \bibinfo{person}{Pengxiang Cheng}, \bibinfo{person}{Zhenhua Dong}, \bibinfo{person}{Xiuqiang He}, \bibinfo{person}{Weike Pan}, {and} \bibinfo{person}{Zhong Ming}.} \bibinfo{year}{2020}\natexlab{}.
\newblock \showarticletitle{A general knowledge distillation framework for counterfactual recommendation via uniform data}. In \bibinfo{booktitle}{\emph{ACM SIGIR Conference on Research and Development in Information Retrieval}}. \bibinfo{pages}{831--840}.
\newblock


\bibitem[Luca(2016)]%
        {luca2016reviews}
\bibfield{author}{\bibinfo{person}{Michael Luca}.} \bibinfo{year}{2016}\natexlab{}.
\newblock \showarticletitle{Reviews, reputation, and revenue: The case of Yelp. com}.
\newblock \bibinfo{journal}{\emph{Com (March 15, 2016). Harvard Business School NOM Unit Working Paper}} \bibinfo{number}{12-016} (\bibinfo{year}{2016}).
\newblock


\bibitem[Meng et~al\mbox{.}(2020)]%
        {meng2020exploring}
\bibfield{author}{\bibinfo{person}{Zaiqiao Meng}, \bibinfo{person}{Richard McCreadie}, \bibinfo{person}{Craig Macdonald}, {and} \bibinfo{person}{Iadh Ounis}.} \bibinfo{year}{2020}\natexlab{}.
\newblock \showarticletitle{Exploring data splitting strategies for the evaluation of recommendation models}. In \bibinfo{booktitle}{\emph{ACM Conference on Recommender Systems}}. \bibinfo{pages}{681--686}.
\newblock


\bibitem[Paterek(2007)]%
        {paterek2007improving}
\bibfield{author}{\bibinfo{person}{Arkadiusz Paterek}.} \bibinfo{year}{2007}\natexlab{}.
\newblock \showarticletitle{Improving regularized singular value decomposition for collaborative filtering}. In \bibinfo{booktitle}{\emph{Proceedings of KDD cup and workshop}}, Vol.~\bibinfo{volume}{2007}. \bibinfo{pages}{5--8}.
\newblock


\bibitem[Pellegrini et~al\mbox{.}(2022)]%
        {pellegrini2022don}
\bibfield{author}{\bibinfo{person}{Roberto Pellegrini}, \bibinfo{person}{Wenjie Zhao}, {and} \bibinfo{person}{Iain Murray}.} \bibinfo{year}{2022}\natexlab{}.
\newblock \showarticletitle{Don’t recommend the obvious: estimate probability ratios}. In \bibinfo{booktitle}{\emph{ACM Conference on Recommender Systems}}. \bibinfo{pages}{188--197}.
\newblock


\bibitem[Schnabel et~al\mbox{.}(2016)]%
        {schnabel2016recommendations}
\bibfield{author}{\bibinfo{person}{Tobias Schnabel}, \bibinfo{person}{Adith Swaminathan}, \bibinfo{person}{Ashudeep Singh}, \bibinfo{person}{Navin Chandak}, {and} \bibinfo{person}{Thorsten Joachims}.} \bibinfo{year}{2016}\natexlab{}.
\newblock \showarticletitle{Recommendations as treatments: Debiasing learning and evaluation}. In \bibinfo{booktitle}{\emph{International Conference on Machine Learning}}. PMLR, \bibinfo{pages}{1670--1679}.
\newblock


\bibitem[Sepliarskaia et~al\mbox{.}(2018)]%
        {sepliarskaia2018preference}
\bibfield{author}{\bibinfo{person}{Anna Sepliarskaia}, \bibinfo{person}{Julia Kiseleva}, \bibinfo{person}{Filip Radlinski}, {and} \bibinfo{person}{Maarten de Rijke}.} \bibinfo{year}{2018}\natexlab{}.
\newblock \showarticletitle{Preference elicitation as an optimization problem}. In \bibinfo{booktitle}{\emph{Proceedings of the 12th ACM Conference on Recommender Systems}}. \bibinfo{pages}{172--180}.
\newblock


\bibitem[Silveira et~al\mbox{.}(2019)]%
        {howGoodIsRecSys}
\bibfield{author}{\bibinfo{person}{Thiago Silveira}, \bibinfo{person}{Min Zhang}, \bibinfo{person}{Xiao Lin}, \bibinfo{person}{Yiqun Liu}, {and} \bibinfo{person}{Shaoping Ma}.} \bibinfo{year}{2019}\natexlab{}.
\newblock \showarticletitle{How good your recommender system is? {A} survey on evaluations in recommendation}.
\newblock \bibinfo{journal}{\emph{Int. J. Mach. Learn. Cybern.}} \bibinfo{volume}{10}, \bibinfo{number}{5} (\bibinfo{year}{2019}), \bibinfo{pages}{813--831}.
\newblock
\urldef\tempurl%
\url{https://doi.org/10.1007/s13042-017-0762-9}
\showDOI{\tempurl}


\bibitem[Steck et~al\mbox{.}(2021)]%
        {steck2021deep}
\bibfield{author}{\bibinfo{person}{Harald Steck}, \bibinfo{person}{Linas Baltrunas}, \bibinfo{person}{Ehtsham Elahi}, \bibinfo{person}{Dawen Liang}, \bibinfo{person}{Yves Raimond}, {and} \bibinfo{person}{Justin Basilico}.} \bibinfo{year}{2021}\natexlab{}.
\newblock \showarticletitle{Deep learning for recommender systems: A Netflix case study}.
\newblock \bibinfo{journal}{\emph{AI Magazine}} \bibinfo{volume}{42}, \bibinfo{number}{3} (\bibinfo{year}{2021}), \bibinfo{pages}{7--18}.
\newblock


\bibitem[Sun(2023)]%
        {sun23newlook}
\bibfield{author}{\bibinfo{person}{Aixin Sun}.} \bibinfo{year}{2023}\natexlab{}.
\newblock \showarticletitle{Take a Fresh Look at Recommender Systems from an Evaluation Standpoint}. In \bibinfo{booktitle}{\emph{Proceedings of the 46th International ACM SIGIR Conference on Research and Development in Information Retrieval}} \emph{(\bibinfo{series}{SIGIR '23})}. \bibinfo{publisher}{ACM}, \bibinfo{address}{New York, NY, USA}, \bibinfo{pages}{2629–2638}.
\newblock
\urldef\tempurl%
\url{https://doi.org/10.1145/3539618.3591931}
\showDOI{\tempurl}


\bibitem[Sun et~al\mbox{.}(2022)]%
        {sun2022daisyrec}
\bibfield{author}{\bibinfo{person}{Zhu Sun}, \bibinfo{person}{Hui Fang}, \bibinfo{person}{Jie Yang}, \bibinfo{person}{Xinghua Qu}, \bibinfo{person}{Hongyang Liu}, \bibinfo{person}{Di Yu}, \bibinfo{person}{Yew-Soon Ong}, {and} \bibinfo{person}{Jie Zhang}.} \bibinfo{year}{2022}\natexlab{}.
\newblock \showarticletitle{DaisyRec 2.0: Benchmarking Recommendation for Rigorous Evaluation}.
\newblock \bibinfo{journal}{\emph{IEEE Transactions on Pattern Analysis and Machine Intelligence}} (\bibinfo{year}{2022}).
\newblock


\bibitem[Sun et~al\mbox{.}(2020)]%
        {sun2020we}
\bibfield{author}{\bibinfo{person}{Zhu Sun}, \bibinfo{person}{Di Yu}, \bibinfo{person}{Hui Fang}, \bibinfo{person}{Jie Yang}, \bibinfo{person}{Xinghua Qu}, \bibinfo{person}{Jie Zhang}, {and} \bibinfo{person}{Cong Geng}.} \bibinfo{year}{2020}\natexlab{}.
\newblock \showarticletitle{Are we evaluating rigorously? benchmarking recommendation for reproducible evaluation and fair comparison}. In \bibinfo{booktitle}{\emph{ACM Conference on Recommender Systems}}. \bibinfo{pages}{23--32}.
\newblock


\bibitem[Tang and Wang(2018)]%
        {tang2018personalized}
\bibfield{author}{\bibinfo{person}{Jiaxi Tang} {and} \bibinfo{person}{Ke Wang}.} \bibinfo{year}{2018}\natexlab{}.
\newblock \showarticletitle{Personalized top-n sequential recommendation via convolutional sequence embedding}. In \bibinfo{booktitle}{\emph{Proceedings of the eleventh ACM international conference on web search and data mining}}. \bibinfo{pages}{565--573}.
\newblock


\bibitem[Woolridge et~al\mbox{.}(2021)]%
        {woolridge2021sequence}
\bibfield{author}{\bibinfo{person}{Daniel Woolridge}, \bibinfo{person}{Sean Wilner}, {and} \bibinfo{person}{Madeleine Glick}.} \bibinfo{year}{2021}\natexlab{}.
\newblock \showarticletitle{Sequence or Pseudo-Sequence? An Analysis of Sequential Recommendation Datasets.}. In \bibinfo{booktitle}{\emph{Perspectives@ RecSys}}.
\newblock


\bibitem[Wu et~al\mbox{.}(2016)]%
        {wu2016collaborative}
\bibfield{author}{\bibinfo{person}{Yao Wu}, \bibinfo{person}{Christopher DuBois}, \bibinfo{person}{Alice~X Zheng}, {and} \bibinfo{person}{Martin Ester}.} \bibinfo{year}{2016}\natexlab{}.
\newblock \showarticletitle{Collaborative denoising auto-encoders for top-n recommender systems}. In \bibinfo{booktitle}{\emph{ACM international conference on web search and data mining}}. \bibinfo{pages}{153--162}.
\newblock


\bibitem[Yu and Sun(2022)]%
        {yusun2022reality}
\bibfield{author}{\bibinfo{person}{Mengying Yu} {and} \bibinfo{person}{Aixin Sun}.} \bibinfo{year}{2022}\natexlab{}.
\newblock \showarticletitle{Dataset vs Reality: Understanding Model Performance from the Perspective of Information Need}.
\newblock \bibinfo{journal}{\emph{CoRR}}  \bibinfo{volume}{abs/2212.02726} (\bibinfo{year}{2022}).
\newblock
\urldef\tempurl%
\url{https://doi.org/10.48550/arXiv.2212.02726}
\showDOI{\tempurl}
\showeprint[arXiv]{2212.02726}


\bibitem[Zangerle and Bauer(2023)]%
        {evaluateRecsysSurvey}
\bibfield{author}{\bibinfo{person}{Eva Zangerle} {and} \bibinfo{person}{Christine Bauer}.} \bibinfo{year}{2023}\natexlab{}.
\newblock \showarticletitle{Evaluating Recommender Systems: Survey and Framework}.
\newblock \bibinfo{journal}{\emph{{ACM} Comput. Surv.}} \bibinfo{volume}{55}, \bibinfo{number}{8} (\bibinfo{year}{2023}), \bibinfo{pages}{170:1--170:38}.
\newblock
\urldef\tempurl%
\url{https://doi.org/10.1145/3556536}
\showDOI{\tempurl}


\bibitem[Zhao et~al\mbox{.}(2022)]%
        {zhao2022revisiting}
\bibfield{author}{\bibinfo{person}{Wayne~Xin Zhao}, \bibinfo{person}{Zihan Lin}, \bibinfo{person}{Zhichao Feng}, \bibinfo{person}{Pengfei Wang}, {and} \bibinfo{person}{Ji-Rong Wen}.} \bibinfo{year}{2022}\natexlab{}.
\newblock \showarticletitle{A revisiting study of appropriate offline evaluation for top-N recommendation algorithms}.
\newblock \bibinfo{journal}{\emph{ACM Transactions on Information Systems}} \bibinfo{volume}{41}, \bibinfo{number}{2} (\bibinfo{year}{2022}), \bibinfo{pages}{1--41}.
\newblock


\bibitem[Zhao et~al\mbox{.}(2021)]%
        {zhao2021recbole}
\bibfield{author}{\bibinfo{person}{Wayne~Xin Zhao}, \bibinfo{person}{Shanlei Mu}, \bibinfo{person}{Yupeng Hou}, \bibinfo{person}{Zihan Lin}, \bibinfo{person}{Yushuo Chen}, \bibinfo{person}{Xingyu Pan}, \bibinfo{person}{Kaiyuan Li}, \bibinfo{person}{Yujie Lu}, \bibinfo{person}{Hui Wang}, \bibinfo{person}{Changxin Tian}, {et~al\mbox{.}}} \bibinfo{year}{2021}\natexlab{}.
\newblock \showarticletitle{Recbole: Towards a unified, comprehensive and efficient framework for recommendation algorithms}. In \bibinfo{booktitle}{\emph{ACM International Conference on Information \& Knowledge Management}}. \bibinfo{pages}{4653--4664}.
\newblock


\end{thebibliography}

\newpage
\appendix
\section{Additional Results}
\label{sec:appdx}

Table~\ref{tab:evaML1M} lists two sets of results of SASRec, TiSASRec, and Caser, on the standard MovieLens dataset. Comparing the results obtained in our experiments with those reported in the original papers, we find that the implementations used in our experiments produce comparable results to those reported in the original papers.

\begin{table}[t]
  \center
   \caption{Results on the standard MovieLens-1M dataset, reported in their original papers, and obtained by using our implementations, respectively.}
  \label{tab:evaML1M}
  \renewcommand\arraystretch{1.2}
  \begin{tabular}{l|c|c|c}
      \toprule
      Model &  Metric &  Paper & Experiment
      \\
      \midrule
      \multirow{2}*{SASRec} & HR@10 & 0.8245 & 0.8238  \\
		~ & NDCG@10 & 0.5905 & 0.5964 \\
      \hline
     \multirow{2}*{TiSASRec} & HR@10 & 0.8038 & 0.8041 \\
      ~ & NDCG@10 & 0.5706 & 0.5697 \\
      \hline
      \multirow{3}*{Caser} & Prec@1 & 0.2502 & 0.2894  \\
		~ & Recall@1 & 0.0148 & 0.0181 \\
     ~ & mAP & 0.1507 & 0.1741 \\
      \hline
  \end{tabular}
\end{table}

\end{document}